\documentclass[12pt]{article}

\usepackage{textcomp}
\usepackage{chetnew}
\usepackage{amssymb,amsmath,amsfonts,mathrsfs}
\usepackage[utf8]{inputenc}
\usepackage{tikz,tikz-cd,url}
\usepackage{xcolor}
\usetikzlibrary{arrows,decorations.markings,shapes.geometric}

\usepackage{bm}
\usepackage[colorinlistoftodos,textsize=tiny]{todonotes}

\renewcommand{\d}[1]{\ensuremath{\operatorname{d}\!{#1}}}

\def\one{{\,\hbox{1\kern-.8mm l}}}

\newcommand{\Tr}{\mathrm{Tr}}

\allowdisplaybreaks
\newcommand{\fp}{\mathfrak{p}}
\newcommand{\fq}{\mathfrak{q}}
\newcommand{\ft}{\mathfrak{t}}
\def\czeta{\vcenter{\hbox{$\scriptstyle\zeta$}}}
\def\zprod#1#2{\ooalign{\hidewidth$#1\czeta$\hidewidth\cr$#1\prod$\cr}}
\DeclareMathOperator*\ZProd{\mathpalette\zprod\relax}

\def\makeatletter{\catcode`\@=11}
\makeatletter
\def\mathbox#1{\hbox{$\m@th#1$}}%
\def\math@ccstyles#1#2#3#4#5#6#7{{\leavevmode
      \setbox0\mathbox{#6#7}%
      \setbox2\mathbox{#4#5}%
      \dimen@ #3%
      \baselineskip\z@\lineskiplimit#1\lineskip\z@
      \vbox{\ialign{##\crcr
             \hfil \kern #2\box2 \hfil\crcr
             \noalign{\kern\dimen@}%
             \hfil\box0\hfil\crcr}}}}
\def\mathaccstyles{\math@ccstyles\maxdimen}
\def\maththroughstyles{\math@ccstyles{-\maxdimen}}
\def\unity%
 {\maththroughstyles{.45\ht0}\z@\displaystyle {\mathchar"006C}\displaystyle 1}

\def\AA{{\cal A}}

\def\HH{{\cal H}}

\def\NN{{\cal N}}
\def\OO{{\cal O}}

\def\RR{{\cal R}}

\def\ZZ{{\cal Z}}

\def\IR{{\mathbb R}}

\def\Tr{{\rm {Tr}}}

\def\d{{\partial}}
\def\p{{\partial}}

\def\beq{\begin{equation}}
\def\eeq{\end{equation}}
\newcommand{\bea}{\begin{eqnarray}}
\newcommand{\eea}{\end{eqnarray}}
\def\bal{\begin{align}}
\def\eal{\end{align}}

\def\noderadius{0.8cm}
\def\dottedlinelength{1cm}

\def\nodecolor{black}
\def\nodeopacity{20}
\def\defnodecolor{blue}
\def\defnodeopacity{20}
\def\deffermicolor{orange}
\def\codimonecolor{red}

\def\arrowstyle{latex}
\def\arrowboundarystyle{>}

\tikzset{
  ->-/.style={
    decoration={
      markings,
      mark=at position #1 with {\arrow{\arrowstyle}}},
    postaction={decorate}
  }
}
\tikzset{
  ->>-/.style={
    dashed,
    decoration={
      markings,
      mark=at position #1 with {\arrow{\arrowboundarystyle\arrowboundarystyle}}},
    postaction={decorate}
  }
}
\tikzset{
  ->>>-/.style={
    dashed,
    decoration={
      markings,
      mark=at position #1 with {\arrow{\arrowboundarystyle\arrowboundarystyle\arrowboundarystyle}}},
    postaction={decorate}
  }
}
\tikzset{square/.style={regular polygon,regular polygon sides=4, inner sep = 0}}

\newcommand{\auxnode}[4]{
  \node (#1) at (#2,#3) {#4};}

\newcommand{\quiverfnode}[4]{
  \node[square, fill=\nodecolor!\nodeopacity, draw, minimum size=\noderadius * 1.41] (#1) at (#2,#3) {#4};}

\newcommand{\quivergnode}[4]{
  \node[circle, fill=\nodecolor!\nodeopacity, draw, minimum size=\noderadius] (#1) at (#2,#3) {#4};}
  
  \newcommand{\defquivergnode}[4]{
  \node[circle, fill=\defnodecolor!\defnodeopacity, draw =\defnodecolor, minimum size=\noderadius] (#1) at (#2,#3) {#4};}

\newcommand{\quiverdefnode}[4]{
  \node (#1) at (#2,#3) {#4};
  \draw[fill = \nodecolor!\nodeopacity] (#2-\noderadius/2 * 1.2, #3 - \noderadius/4) -- (#2 + \noderadius/2 * 1.2, #3 - \noderadius/4) arc (0:180:\noderadius/2 * 1.2);
  \node (#1) at (#2,#3) {#4}; }
  
\newcommand{\defverbifline}[2]{
  \draw[->-=0.6, color = \defnodecolor] ([xshift=1mm] #1.center) to ([xshift=1mm] #2.center);
  \draw[->-=0.6, color = \defnodecolor] ([xshift=-1mm] #2.center) to ([xshift=-1mm] #1.center);}  
  
\newcommand{\fhorbifline}[3]{
  \draw[->-=#1,color = \deffermicolor] ([yshift=1mm] #2.center) to ([yshift=1mm] #3.center);
  \draw[->-=#1, color = \defnodecolor] ([yshift=-1mm] #2.center) to ([yshift=-1mm] #3.center);}

\newcommand{\lenfermline}[3]{
  \draw[->-=#1, color = \deffermicolor] (#2) to (#3);}

\newcommand{\quiverline}[2]{
  \draw[->-=0.65] (#1) to (#2);}
  
\newcommand{\lenquiverline}[3]{
  \draw[->-=#1] (#2) to (#3);}
  
 \newcommand{\defquiverline}[3]{
  \draw[->-= #1, color = \defnodecolor] (#2) to (#3);}
  
 \newcommand{\sdefquiverline}[2]{
  \draw[->-=1cm, color = \defnodecolor] (#1) to (#2);}
  
\newcommand{\quiverdottedline}[2]{
  \draw[dotted] (#1) to ++(#2:\dottedlinelength);}
  
\newcommand{\defquiverdottedline}[2]{
  \draw[dotted, color = \defnodecolor] (#1) to ++(#2:\dottedlinelength);}
  
\newcommand{\ferquiverdottedline}[2]{
  \draw[dotted, color = \deffermicolor] (#1) to ++(#2:\dottedlinelength);}
 
  \newcommand{\douquiverdottedline}[2]{
  \draw[dotted, color = \deffermicolor] ([yshift=1mm] #1.center) to ++(#2:\dottedlinelength);
  \draw[dotted, color = \defnodecolor] ([yshift=-1mm] #1.center) to ++(#2:\dottedlinelength);}

\newcommand{\bedefadjointlink}[1]{
 \draw[->- = 1.4cm,color = \defnodecolor] (#1.270) arc (90:450:4mm);}

\newcommand{\hypquiverline}[2]{
  \draw (#1) to (#2);}

\newcommand{\abhypadjointlink}[1]{
 \draw (#1.90) arc (-90:270:4mm);}

\newcommand{\codimonenode}[4]{
  \node (#1) at (#2,#3) {#4};
  \draw[draw =  \codimonecolor, fill = \codimonecolor!\nodeopacity] (#2-\noderadius/2 * 1.2, #3 - \noderadius/4) -- (#2 + \noderadius/2 * 1.2, #3 - \noderadius/4) arc (0:180:\noderadius/2 * 1.2);
  \node (#1) at (#2,#3) {#4}; }

\newcommand{\codimonebifline}[2]{
  \draw[->-=0.6, color =  \codimonecolor] ([xshift=1mm] #1.center) to ([xshift=1mm] #2.center);
  \draw[->-=0.6, color =  \codimonecolor] ([xshift=-1mm] #2.center) to ([xshift=-1mm] #1.center);}
  
\newcommand{\horcodimonebifline}[2]{
  \draw[->-=0.6,  color = \codimonecolor] ([yshift=1mm] #1.center) to ([yshift=1mm] #2.center);
  \draw[->-=0.6] ([yshift=-1mm] #1.center) to ([yshift=-1mm] #2.center);}

\def\Ihnodedist{1.5cm}
\def\Ivnodedist{1.5cm}
\def\Ilastnode{6.5cm}

\newcommand{\drawquiverpuresym}{
\begin{tikzpicture}[thick]
  \auxnode{A01}{-\Ihnodedist}{0}{}
  \auxnode{A11}{0}{0}{}
  \auxnode{A21}{\Ihnodedist}{0}{}
  \auxnode{A31}{\Ihnodedist * 2}{0}{}
  \auxnode{A41}{\Ilastnode }{0}{}
  \auxnode{Am1}{\Ilastnode +\Ihnodedist}{0}{}
  
	\quiverline{A01}{A11}
	\quiverline{A11}{A21}  
	\quiverline{A21}{A31}  
	\quiverline{A41}{Am1}      
  
	\quivergnode{G11}{0}{0}{{\small $k$}}  
	\quivergnode{G21}{\Ihnodedist}{0}{{\small $k$}}  
	\quivergnode{G31}{\Ihnodedist * 2}{0}{{\small $k$}} 
	\quivergnode{G41}{\Ilastnode}{0}{{\small $k$}}   
	
	\quiverdottedline{G31}{0}
	\quiverdottedline{G41}{180}
  
  \draw[dashed] (-\Ihnodedist/2, -\Ivnodedist/2) to (-\Ihnodedist/2, \Ivnodedist/2);
  \draw[dashed] (\Ilastnode + \Ihnodedist/2, -\Ivnodedist/2) to (\Ilastnode+ \Ihnodedist/2, \Ivnodedist/2);
  
	\draw [draw = none,fill = white] (-\Ihnodedist/2, -\Ivnodedist/2) rectangle (-\Ihnodedist, \Ivnodedist/2);
	\draw [draw = none,fill = white] (\Ilastnode +\Ihnodedist/2, -\Ivnodedist/2) rectangle (\Ilastnode +\Ihnodedist, \Ivnodedist/2);
  
\end{tikzpicture}
}

\def\IIhnodedist{1.5cm}
\def\IIvnodedist{1.5cm}
\def\IIlastnode{6.5cm}

\newcommand{\drawquiversqcd}{
\begin{tikzpicture}[thick]
  \auxnode{A01}{-\IIhnodedist}{0}{}
  \auxnode{A11}{0}{0}{}
  \auxnode{A21}{\IIhnodedist}{0}{}
  \auxnode{A31}{\IIhnodedist * 2}{0}{}
  \auxnode{A41}{\IIlastnode }{0}{}
  \auxnode{Am1}{\IIlastnode +\IIhnodedist}{0}{}
  
  \auxnode{AF01}{-\IIhnodedist}{\IIvnodedist}{}
  \auxnode{AF11}{0}{\IIvnodedist}{}
  \auxnode{AF21}{\IIhnodedist}{\IIvnodedist}{}
  \auxnode{AF31}{\IIhnodedist * 2}{\IIvnodedist}{}
  \auxnode{AF41}{\IIlastnode }{\IIvnodedist}{}

	\quiverline{A01}{A11}
	\quiverline{A11}{A21}  
	\quiverline{A21}{A31}  
	\quiverline{A41}{Am1}   
	 
	\quiverline{AF11}{A11}
	\quiverline{AF21}{A21}  
	\quiverline{AF31}{A31}  
	\quiverline{AF41}{Am1}   
	
	\quiverline{A11}{AF01}
	\quiverline{A21}{AF11}  
	\quiverline{A31}{AF21}  
	\quiverline{AF41}{A41}   
	
	\quiverdottedline{G31}{0}
	\quiverdottedline{G41}{180}
	\quiverdottedline{AF31}{-45}

	\quivergnode{G11}{0}{0}{{\small $k$}}  
	\quivergnode{G21}{\IIhnodedist}{0}{{\small $k$}}  
	\quivergnode{G31}{\IIhnodedist * 2}{0}{{\small $k$}} 
	\quivergnode{G41}{\IIlastnode}{0}{{\small $k$}}  
	
	\quiverfnode{F11}{0}{\IIvnodedist}{{\small $N_f$}}
	\quiverfnode{F21}{\IIhnodedist}{\IIvnodedist}{{\small $N_f$}}
    \quiverfnode{F31}{\IIhnodedist * 2}{\IIvnodedist}{{\small $N_f$}}
	\quiverfnode{F41}{\IIlastnode }{\IIvnodedist}{{\small $N_f$}}

  \draw[dashed] (-\Ihnodedist/2, -\Ivnodedist/2) to (-\Ihnodedist/2, \Ivnodedist * 3/2);
  \draw[dashed] (\Ilastnode + \Ihnodedist/2, -\Ivnodedist/2) to (\Ilastnode+ \Ihnodedist/2, \Ivnodedist * 3/2);
  
	\draw [draw = none,fill = white] (-\Ihnodedist/2, -\Ivnodedist/2) rectangle (-\Ihnodedist, \Ivnodedist * 3/2);
	\draw [draw = none,fill = white] (\Ilastnode +\Ihnodedist/2, -\Ivnodedist/2) rectangle (\Ilastnode +\Ihnodedist, \Ivnodedist * 3/2);
  
\end{tikzpicture}
}
 
\def\IIIhnodedist{3cm}
\def\IIIvnodedist{1.5cm}
\def\IIIlastnode{10.5cm}

\newcommand{\threeDdefquivone}{
\begin{tikzpicture}[thick]
	\auxnode{AG01}{-\IIIhnodedist/2}{0}{}
	\auxnode{AG11}{0}{0}{}
	\auxnode{AG21}{\IIIhnodedist}{0}{}
	\auxnode{AG31}{\IIIhnodedist * 2}{0}{}
	\auxnode{AG41}{\IIIlastnode}{0}{}
	\auxnode{AGm1}{\IIIlastnode +\IIIhnodedist}{0}{}

	\auxnode{AD00}{-\IIIhnodedist/2}{-\IIIvnodedist}{}
	\auxnode{AD10}{\IIIhnodedist/2}{-\IIIvnodedist}{}
	\auxnode{AD20}{\IIIhnodedist *3/2}{-\IIIvnodedist}{}
	\auxnode{AD30}{\IIIhnodedist *5/2}{-\IIIvnodedist}{}
	\auxnode{AD40}{\IIIlastnode+\IIIhnodedist/2}{-\IIIvnodedist}{}
	\auxnode{ADm0}{\IIIlastnode+\IIIhnodedist }{-\IIIvnodedist}{}
	
	\draw[loosely dashed,color= \nodecolor!30 ] (-\IIIhnodedist/2,-\IIIvnodedist/2) -- (\IIIlastnode + \IIIhnodedist ,-\IIIvnodedist/2);

	\hypquiverline{AG01}{AG11}
	\hypquiverline{AG11}{AG21}
	\hypquiverline{AG21}{AG31}
	\hypquiverline{AGm1}{AG41}
	
	\sdefquiverline{AG11}{AD00}
	\sdefquiverline{AD10}{AG11}
	\sdefquiverline{AG21}{AD10}
	\sdefquiverline{AD20}{AG21}
	\sdefquiverline{AG31}{AD20}
	\sdefquiverline{AD30}{AG31}
	\sdefquiverline{AGm1}{AD40}
	\sdefquiverline{AD40}{AG41}

	\defquiverline{\IIIhnodedist/2}{AD00}{AD10}
	\defquiverline{\IIIhnodedist/2}{AD10}{AD20}
	\defquiverline{\IIIhnodedist/2}{AD20}{AD30}
	\defquiverline{\IIIhnodedist/2}{AD40}{ADm0}

	\defquiverdottedline{AD30}{45}
	\defquiverdottedline{AD30}{0}
	\defquiverdottedline{AD40}{180}
	\quiverdottedline{AG31}{0}
	\quiverdottedline{AG41}{180}
	\defquiverdottedline{AG41}{225}
	
	\quiverdefnode{G11}{0}{0}{{\small $k$}}
	\quiverdefnode{G21}{\IIIhnodedist}{0}{{\small $k$}}
	\quiverdefnode{G31}{\IIIhnodedist * 2}{0}{{\small $k$}}
	\quiverdefnode{G41}{\IIIlastnode}{0}{{\small $k$}}

	\defquivergnode{D10}{\IIIhnodedist/2}{-\IIIvnodedist}{{\small \textcolor{blue}{$n$}}}
	\defquivergnode{D20}{\IIIhnodedist *3/2}{-\IIIvnodedist}{{\small \textcolor{blue}{$n$}}}
	\defquivergnode{D30}{\IIIhnodedist *5/2}{-\IIIvnodedist}{{\small \textcolor{blue}{$n$}}}
	\defquivergnode{D40}{\IIIlastnode+\IIIhnodedist/2}{-\IIIvnodedist}{{\small \textcolor{blue}{$n$}}}

	\draw[dashed] (-\IIIhnodedist/4,-\IIIvnodedist *3/2) to 						(-\IIIhnodedist/4,\IIIvnodedist  *1/2);
	\draw[dashed] (\IIIlastnode + \IIIhnodedist* 3/4,-\IIIvnodedist *3/2) to (\IIIlastnode + \IIIhnodedist* 3/4,\IIIvnodedist  *1/2);
  
	\draw [draw = none,fill = white] (-\IIIhnodedist/4 - 0.35,-\IIIvnodedist *3/2) rectangle (-\IIIhnodedist/2-0.5cm,\IIIvnodedist  *1/2);
	\draw [draw = none,fill = white] (\IIIlastnode +\IIIhnodedist* 3/4 + 						0.35,-\IIIvnodedist *3/2) rectangle (\IIIlastnode 							+\IIIhnodedist + 	0.5cm,\IIIvnodedist  *1/2);

	\node at (-\IIIhnodedist * 1/2,-\IIIvnodedist/4) {3D};
	\node at (-\IIIhnodedist * 1/2,-\IIIvnodedist* 3/4) {1D};

\end{tikzpicture}
}

\def\IVhnodedist{1.5cm}
\def\IVvnodedist{1.5cm}

\newcommand{\fourDdefquivone}{
\begin{tikzpicture}[thick]
	\auxnode{AG11}{0}{0}{}
	
	\auxnode{AD10}{0}{-\IVvnodedist}{}

	\draw[loosely dashed,color= \nodecolor!30 ] (-\IVhnodedist,-\IVvnodedist/2) -- ( \IVhnodedist,-\IVvnodedist/2);
 
	\node at (-\IVhnodedist * 3/4,-\IVvnodedist/4) {4D};
	\node at (-\IVhnodedist * 3/4,-\IVvnodedist* 3/4) {2D};	
		
	\defverbifline{AG11}{AD10}

	\bedefadjointlink{AD10}

	\abhypadjointlink{AG11}

	\quiverdefnode{G11}{0}{0}{{\small $k$}}
	
	\defquivergnode{D10}{0}{-\IVvnodedist}{{\small \textcolor{blue}{$n$}}}
	
\end{tikzpicture}
}

\def\Vhnodedist{3cm}
\def\Vvnodedist{1.5cm}
\def\Vlastnode{10.5cm}
\def\Vdefnodemult{2}

\newcommand{\drawquiverdefextwo}{
\begin{tikzpicture}[thick]

	\auxnode{AG01}{-\Vhnodedist}{0}{}
	\auxnode{AG11}{0}{0}{}
	\auxnode{AG21}{\Vhnodedist}{0}{}
	\auxnode{AG31}{\Vhnodedist * 2}{0}{}
	\auxnode{AG41}{\Vlastnode}{0}{}
	\auxnode{AG51}{\Vlastnode +\Vhnodedist}{0}{}

	\auxnode{AF00}{-\Vhnodedist}{-\Vvnodedist}{}
	\auxnode{AF10}{0}{-\Vvnodedist}{}
	\auxnode{AF20}{\Vhnodedist}{-\Vvnodedist}{}
	\auxnode{AF30}{\Vhnodedist *2 }{-\Vvnodedist}{}
	\auxnode{AF40}{\Vlastnode }{-\Vvnodedist}{}
	\auxnode{AF50}{\Vlastnode+\Vhnodedist }{-\Vvnodedist}{}

	\auxnode{AF02}{-\Vhnodedist}{\Vvnodedist}{}
	\auxnode{AF12}{0}{\Vvnodedist}{}
	\auxnode{AF22}{\Vhnodedist}{\Vvnodedist}{}
	\auxnode{AF32}{\Vhnodedist *2 }{\Vvnodedist}{}
	\auxnode{AF42}{\Vlastnode }{\Vvnodedist}{}
	\auxnode{AF52}{\Vlastnode+\Vhnodedist }{\Vvnodedist}{}	
	
	\auxnode{AD00}{-\Vhnodedist /2}{-\Vvnodedist *\Vdefnodemult}{}
	\auxnode{AD10}{\Vhnodedist /2}{-\Vvnodedist*\Vdefnodemult}{}
	\auxnode{AD20}{\Vhnodedist *3/2}{-\Vvnodedist*\Vdefnodemult}{}
	\auxnode{AD30}{\Vhnodedist *5/2}{-\Vvnodedist*\Vdefnodemult}{}
	\auxnode{AD40}{\Vlastnode + \Vhnodedist /2}{-\Vvnodedist*\Vdefnodemult}{}
	\auxnode{AD50}{\Vlastnode + \Vhnodedist *3/2 }{-\Vvnodedist*\Vdefnodemult}{}

	\draw[loosely dashed,color= \nodecolor!30 ] (-\Vhnodedist/2, -\Vvnodedist*3/2) -- (\Vlastnode + \Vhnodedist , -\Vvnodedist *3/2);

	\lenquiverline{\Vhnodedist/2}{AG01}{AG11}
	\lenquiverline{\Vhnodedist/2}{AG11}{AG21}
	\lenquiverline{\Vhnodedist/2}{AG21}{AG31}
	\lenquiverline{\Vhnodedist/2}{AG41}{AG51}
	
	\lenquiverline{\Vvnodedist/2}{AF12}{AG11}
	\lenquiverline{\Vvnodedist/2}{AF22}{AG21}
	\lenquiverline{\Vvnodedist/2}{AF32}{AG31}
	\lenquiverline{\Vvnodedist/2}{AF42}{AG41}
	
	\lenquiverline{\Vvnodedist/2}{AG11}{AF10}
	\lenquiverline{\Vvnodedist/2}{AG21}{AF20}
	\lenquiverline{\Vvnodedist/2}{AG31}{AF30}
	\lenquiverline{\Vvnodedist/2}{AG41}{AF40}

	\lenquiverline{\Vvnodedist }{AG11}{AF02}
	\lenquiverline{\Vvnodedist }{AG21}{AF12}
	\lenquiverline{\Vvnodedist }{AG31}{AF22}
	\lenquiverline{\Vvnodedist }{AG51}{AF42}
	
	\lenquiverline{\Vvnodedist }{AF10}{AG01}
	\lenquiverline{\Vvnodedist }{AF20}{AG11}
	\lenquiverline{\Vvnodedist }{AF30}{AG21}
	\lenquiverline{\Vvnodedist }{AF50}{AG41}

	\fhorbifline{\Vhnodedist/2}{AD00}{AD10}
	\fhorbifline{\Vhnodedist/2}{AD10}{AD20}
	\fhorbifline{\Vhnodedist/2}{AD20}{AD30}
	\fhorbifline{\Vhnodedist/2}{AD40}{AD50}
	
	\lenfermline{\Vhnodedist/2}{AD10}{AG11}
	\lenfermline{\Vhnodedist/2}{AD20}{AG21}
	\lenfermline{\Vhnodedist/2}{AD30}{AG31}
	\lenfermline{\Vhnodedist/2}{AD40}{AG41}
	
	\lenfermline{\Vhnodedist/4}{AF10}{AD00}
	\lenfermline{\Vhnodedist/4}{AF20}{AD10}
	\lenfermline{\Vhnodedist/4}{AF30}{AD20}
	\lenfermline{\Vhnodedist/4}{AF50}{AD40}

	\defquiverline{\Vhnodedist/2}{AD00}{AG11}
	\defquiverline{\Vhnodedist/2}{AD10}{AG21}
	\defquiverline{\Vhnodedist/2}{AD20}{AG31}
	\defquiverline{\Vhnodedist/2}{AD40}{AG51}

	\defquiverline{\Vhnodedist/4}{AF10}{AD10}
	\defquiverline{\Vhnodedist/4}{AF20}{AD20}
	\defquiverline{\Vhnodedist/4}{AF30}{AD30}
	\defquiverline{\Vhnodedist/4}{AF40}{AD40}

	\bedefadjointlink{AD10}
	\bedefadjointlink{AD20}
	\bedefadjointlink{AD30}
	\bedefadjointlink{AD40}
	
	\quiverdottedline{AF32}{-30}
	\quiverdottedline{AG31}{0}
	\quiverdottedline{AG31}{-30}
	\defquiverdottedline{AG41}{240}
	\quiverdottedline{AG41}{180}
	\quiverdottedline{AG41}{150}
	\quiverdottedline{AF40}{150}
	\ferquiverdottedline{AF40}{210}
	\defquiverdottedline{AD30}{60}
	\ferquiverdottedline{AD30}{30}
	\douquiverdottedline{AD30}{0}
	\douquiverdottedline{AD40}{180}
	
	\quiverdefnode{G11}{0}{0}{{\small $k$}}
	\quiverdefnode{G21}{\Vhnodedist}{0}{{\small $k$}}
	\quiverdefnode{G31}{\Vhnodedist * 2}{0}{{\small $k$}}
	\quiverdefnode{G41}{\Vlastnode}{0}{{\small $k$}}

	\quiverfnode{F10}{0}{-\Vvnodedist}{{\small $N_2$}}
	\quiverfnode{F20}{\Vhnodedist}{-\Vvnodedist}{{\small $N_2$}}
	\quiverfnode{F30}{\Vhnodedist *2 }{-\Vvnodedist}{{\small $N_2$}}
	\quiverfnode{F40}{\Vlastnode }{-\Vvnodedist}{{\small $N_2$}}
	\quiverfnode{F12}{0}{\Vvnodedist}{{\small $N_1$}}
	\quiverfnode{F22}{\Vhnodedist}{\Vvnodedist}{{\small $N_1$}}
	\quiverfnode{F32}{\Vhnodedist *2 }{\Vvnodedist}{{\small $N_1$}}
	\quiverfnode{F42}{\Vlastnode }{\Vvnodedist}{{\small $N_1$}}
	
	\defquivergnode{D10}{\Vhnodedist /2}{-\Vvnodedist *\Vdefnodemult}{{\small \textcolor{blue}{$n$}}}
	\defquivergnode{D20}{\Vhnodedist *3/2}{-\Vvnodedist *\Vdefnodemult}{{\small \textcolor{blue}{$n$}}}
	\defquivergnode{D30}{\Vhnodedist *5/2}{-\Vvnodedist *\Vdefnodemult}{{\small \textcolor{blue}{$n$}}}
	\defquivergnode{D40}{\Vlastnode + \Vhnodedist /2}{-\Vvnodedist *\Vdefnodemult}{{\small \textcolor{blue}{$n$}}}
	
	\draw[dashed] (-\Vhnodedist/4,-\Vvnodedist*1/2 -\Vvnodedist *\Vdefnodemult) to 						(-\Vhnodedist/4,\Vvnodedist  *3/2);
	\draw[dashed] (\Vlastnode + \Vhnodedist* 3/4,-\Vvnodedist*1/2 -\Vvnodedist *\Vdefnodemult) to (\Vlastnode + \Vhnodedist* 3/4,\Vvnodedist  *3/2);
  
	\draw [draw = none,fill = white] (-\Vhnodedist/4 - 0.35,-\Vvnodedist*1/2 -\Vvnodedist *\Vdefnodemult) rectangle (-\Vhnodedist -0.5cm,\Vvnodedist  *3/2);
	\draw [draw = none,fill = white] (\Vlastnode +\Vhnodedist* 3/4 + 						0.35,-\Vvnodedist*1/2 -\Vvnodedist *\Vdefnodemult) rectangle (\Vlastnode 							+\Vhnodedist *3/2 + 	0.5cm,\Vvnodedist  *3/2);

	\node at (-\Vhnodedist * 1/2,-\Vvnodedist *3/2+\Vvnodedist/4) {3D};
	\node at (-\Vhnodedist * 1/2,-\Vvnodedist *3/2-\Vvnodedist/4) {1D};

\end{tikzpicture}
}

\def\VIIhnodedist{1.5cm}
\def\VIIvnodedist{1.5cm}

\newcommand{\drawquiverdefexthreeres}{
\begin{tikzpicture}[thick]
	\auxnode{AF11}{0}{0}{}
	\auxnode{AG21}{\VIIhnodedist}{0}{}
	\auxnode{AF31}{\VIIhnodedist *2}{0}{}
	
	\auxnode{AD20}{\VIIhnodedist}{-\VIIvnodedist}{}
	
	\draw[loosely dashed,color= \nodecolor!30 ] (-\VIIhnodedist,-\VIIvnodedist/2) -- ( \VIIhnodedist *3,-\VIIvnodedist/2);
 
	\node at (-\VIIhnodedist * 3/4,-\VIIvnodedist/4) {4D};
	\node at (-\VIIhnodedist * 3/4,-\VIIvnodedist* 3/4) {2D};	
		
	\hypquiverline{AF11}{AG21}
	\hypquiverline{AG21}{AF31}
	
	\defquiverline{\VIIvnodedist/2}{AD20}{AG21}
	\sdefquiverline{AF31}{AD20}
	
	\bedefadjointlink{AD20}
	
	\quiverfnode{F11}{0}{0}{{\small $N_1$}}
	\quiverdefnode{G21}{\VIIhnodedist}{0}{{\small $k$}}
	\quiverfnode{F31}{\VIIhnodedist *2}{0}{{\small $N_2$}}
	
	\defquivergnode{D20}{\VIIhnodedist}{-\VIIvnodedist}{{\small \textcolor{blue}{$n$}}}
	
\end{tikzpicture}
}

\def\Xhnodedist{1.5cm}
\def\Xvnodedist{1.5cm}
\def\Xlastnode{6.5cm}

\newcommand{\drawquivercodimonea}{
\begin{tikzpicture}[thick]
  \auxnode{A01}{-\Xhnodedist}{0}{}
  \auxnode{A11}{0}{0}{}
  \auxnode{A21}{\Xhnodedist}{0}{}
  \auxnode{A31}{\Xhnodedist * 2}{0}{}
  \auxnode{A41}{\Xlastnode }{0}{}
  \auxnode{Am1}{\Xlastnode +\Xhnodedist}{0}{}
  
	\quiverline{A01}{A11}
	\quiverline{A11}{A21}  
	\horcodimonebifline{A21}{A31}  
	\quiverline{A41}{Am1}      
  
	\quivergnode{G11}{0}{0}{{\small $k$}}  
	\quivergnode{G21}{\Xhnodedist}{0}{{\small $k$}}  
	\quivergnode{G31}{\Xhnodedist * 2}{0}{{\small $k$}} 
	\quivergnode{G41}{\Xlastnode}{0}{{\small $k$}}   
	
	\quiverdottedline{G31}{0}
	\quiverdottedline{G41}{180}
  
  \draw[dashed] (-\Xhnodedist/2, -\Xvnodedist/2) to (-\Xhnodedist/2, \Xvnodedist/2);
  \draw[dashed] (\Xlastnode + \Xhnodedist/2, -\Xvnodedist/2) to (\Xlastnode+ \Xhnodedist/2, \Xvnodedist/2);
  
	\draw [draw = none,fill = white] (-\Xhnodedist/2, -\Xvnodedist/2) rectangle (-\Xhnodedist, \Xvnodedist/2);
	\draw [draw = none,fill = white] (\Xlastnode +\Xhnodedist/2, -\Xvnodedist/2) rectangle (\Xlastnode +\Xhnodedist, \Xvnodedist/2);
  
\end{tikzpicture}
}

\def\Xhnodedist{1.5cm}
\def\Xvnodedist{1.5cm}
\def\Xlastnode{6.5cm}

\newcommand{\drawquivercodimoneb}{
\begin{tikzpicture}[thick]
  \auxnode{A01}{-\Xhnodedist}{0}{}
  \auxnode{A11}{0}{0}{}
  \auxnode{A21}{\Xhnodedist}{0}{}
  \auxnode{A31}{\Xhnodedist * 2}{0}{}
  \auxnode{A41}{\Xlastnode }{0}{}
  \auxnode{Am1}{\Xlastnode +\Xhnodedist}{0}{}
  \auxnode{A30}{\Xhnodedist *2}{-\Xvnodedist}{}
  
	\quiverline{A01}{A11}
	\quiverline{A11}{A21}  
	\quiverline{A21}{A31}  
	\quiverline{A41}{Am1}      

	\codimonebifline{A31}{A30}

	\quivergnode{G11}{0}{0}{{\small $k$}}  
	\quivergnode{G21}{\Xhnodedist}{0}{{\small $k$}}  
	\quivergnode{G31}{\Xhnodedist * 2}{0}{{\small $k$}} 
	\quivergnode{G41}{\Xlastnode}{0}{{\small $k$}}   
	
	\codimonenode{D30}{\Xhnodedist *2}{-\Xvnodedist}{{\small \textcolor{red}{$k^\prime$}}}
	
	\quiverdottedline{G31}{0}
	\quiverdottedline{G41}{180}
  
  \draw[dashed] (-\Xhnodedist/2, -\Xvnodedist/2) to (-\Xhnodedist/2, \Xvnodedist/2);
  \draw[dashed] (\Xlastnode + \Xhnodedist/2, -\Xvnodedist/2) to (\Xlastnode+ \Xhnodedist/2, \Xvnodedist/2);
  
	\draw [draw = none,fill = white] (-\Xhnodedist/2, -\Xvnodedist/2) rectangle (-\Xhnodedist, \Xvnodedist/2);
	\draw [draw = none,fill = white] (\Xlastnode +\Xhnodedist/2, -\Xvnodedist/2) rectangle (\Xlastnode +\Xhnodedist, \Xvnodedist/2);
  
\end{tikzpicture}
}

\preprint{QMUL-PH-18-24 \\ DCPT-18-31}

\title{Deconstructing  Defects}

\author{Joseph Hayling,$^{a,\heartsuit}$ Vasilis Niarchos,$^{b,\clubsuit}$ and Constantinos Papageorgakis\;$^{a,\diamondsuit}$}

\affiliation{$^a$CRST and School of Physics and Astronomy\\ Queen Mary University of London, London E1 4NS, UK\\ $ $ \\ $^b$Department of Mathematical Sciences and Centre for Particle Theory\\ Durham University, Durham DH1 3LE, UK

\emails{$^\heartsuit$j.a.hayling@qmul.ac.uk, $^\clubsuit$vasileios.niarchos@durham.ac.uk, $^\diamondsuit$c.papageorgakis@qmul.ac.uk}
}

\abstract{We use the exact-deconstruction prescription to lift various squashed-$S^3$ partition functions with supersymmetric-defect insertions to four-dimensional superconformal indices. Starting from three-dimensional circular-quiver theories with vortex-loop-operator insertions, we recover the index of four-dimensional theories in the presence of codimension-two surface defects with (2,2) supersymmetry. The case of deconstruction with Wilson-loop insertions is discussed separately. We provide evidence that a certain prescription leads to the index of four-dimensional theories in the presence of surface defects with (4,0) supersymmetry. In addition, we deconstruct the index of four-dimensional gauge theories with codimension-one $\frac{1}{2}$-BPS defects, starting from three-dimensional circular-quiver theories containing localised matter/gauge-field insertions at specific nodes. We also clarify certain calculational and conceptual points related to exact deconstruction.}

\date{\today}

\begin{document}

\maketitle

\hypersetup{pageanchor=true}

\setcounter{tocdepth}{2}

\toc

\section{Introduction and Summary}\label{intro}

Over the last decade, there has been remarkable progress in the calculation of supersymmetric partition functions. A host of examples in diverse dimensions can be calculated exactly, i.e.\ to all orders in the coupling, using powerful techniques such as the superconformal index and supersymmetric localisation \cite{Kinney:2005ej,Romelsberger:2005eg,Pestun:2007rz,Pestun:2016zxk}. Somewhat more recently, this success has been extended to include contributions of supersymmetric defects of different codimensionalities \cite{Gang:2012ff,Gaiotto:2012xa,Gadde:2013ftv,Drukker:2012sr,Assel:2015oxa,Gomis:2016ljm,Cordova:2017ohl}. The results of the above calculations can often be expressed elegantly in terms of special functions. This reformulation sheds light into various properties of the theory itself, like its duality structure. Furthermore, mathematical relationships between different special functions can lead to connections between supersymmetric partition functions in different dimensions via dimensional reduction \cite{Dolan:2011rp,Gadde:2011ia,Niarchos:2012ah,Aharony:2013dha}.\footnote{See also \cite{Amariti:2016kat} for a related review.}

In this paper we explore the above themes in a process which can be thought of as the reverse of dimensional reduction, namely dimensional deconstruction \cite{ArkaniHamed:2001ca,Hill:2000mu}. In its simplest formulation, dimensional deconstruction involves starting with a circular-quiver gauge theory and employing a finely-tuned infinite-node limit to obtain dynamics for a theory in one-compact dimension higher. Although in its original form deconstruction relates four-dimensional quivers to five-dimensional theories on a circle, it is possible under certain conditions to deconstruct theories in any dimension starting from a lower-dimensional compact quiver lattice\cite{Kaplan:2002wv}.\footnote{Recent interesting work on dimensional deconstruction includes \cite{Bourget:2017sxr,Aitken:2018joz}. For an application to the six-dimensional (2,0) theory and Little String Theory see \cite{ArkaniHamed:2001ie,Lambert:2012qy}.} Here we will be interested in the version that relates three-dimensional quivers to four-dimensional theories on a circle. Moreover, we will mainly focus on examples that contain supersymmetric defects. Defects have not been considered previously in the context of deconstruction. 

It should be noted that the main argument behind the dimensional-deconstruction proposal uses the Lagrangian description of the quiver to identify the classical Kaluza--Klein (KK) spectrum for the higher-dimensional theory. However, one can actually go beyond this perturbative approach and apply the principle of deconstruction at the level of full supersymmetric partition functions on compact manifolds \cite{Hayling:2017cva,Hayling:2018fmv}. We will refer to this operation as ``exact deconstruction'' to distinguish it from the dimensional-deconstruction limit of \cite{ArkaniHamed:2001ca,Hill:2000mu}. One of the purposes of this paper is to augment the known applications of exact deconstruction and clarify aspects of the operations it involves.

In the main part of this article we will apply exact deconstruction to the squashed-$S^3$ partition functions\footnote{We will focus on squashings of the $S^3$ that preserve a $U(1)\times U(1)$ isometry.} of three-dimensional circular-quiver theories in the presence of various defects, to recover the $S^3 \times S^1$-partition function---also known as the index---of four-dimensional theories that include $\frac{1}{2}$-BPS defects \emph{wrapping} the emerging circle. We will also deconstruct BPS defects \emph{localised} on the emerging circle.

A more thorough list of our main results involves the following points:
\begin{itemize}
\item
  We recover the 4D-2D index of four-dimensional $\mathcal N=2$ theories in the presence of  codimension-two defects with \emph{nonchiral} $(2,2)$ supersymmetry \cite{Gadde:2013ftv}, starting from 3D-1D sphere partition functions of three-dimensional quivers including supersymmetric vortex loops \cite{Drukker:2012sr,Assel:2015oxa}. In order to achieve this, we need to implement the exact-deconstruction procedure on the coupled 3D-1D partition function and then recast the result using two key mathematical identities between special functions. For the 3D$\to$4D part we employ the following relation between hyperbolic Gamma functions (which appear in the expression of three-sphere partition functions) and elliptic Gamma functions (which appear in the expression of four-dimensional indices)
  \begin{align}
     \nonumber
\prod_{\alpha = -\infty}^\infty \Gamma_h\left(x + \frac{\alpha}{R}\bigg| \omega_1 , \omega_2\right) = \mathfrak x^2 \left(\fp \fq\right)^{-\frac{1}{2}} \Gamma_e (\mathfrak x |\fp, \fq)\;,
  \end{align}
where $\mathfrak x=e^{2\pi i R x}$, $\fp = e^{2\pi i R \omega_1}$, $\fq= e^{2\pi i R \omega_2}$ and $R$ is a dimensionless quantity proportional to the radius of the deconstructed circle. For the 1D$\to$2D part we use an analogous identity relating the special functions that appear in the 1D partition function, $\Delta_h$, to the ones appearing in the 2D elliptic genus, $\Delta_e$,
  \begin{align}\nonumber
\prod_{\alpha=-\infty}^\infty \Delta_h \left( a + i\frac{\alpha}{R} \Big | \omega, t \right) = \Delta_e(A | \Omega,T) ~,
  \end{align}
where $A= e^{-2\pi R a}$, $\Omega= e^{-2\pi  R \omega }$, $T = e^{-2\pi R t}$. Details pertaining to these functions are provided in the main text and summarised in Appendix \ref{math}.    

\item We propose expressions for the superconformal index of 4D $\NN = 2$ theories in the presence of codimension-two defects with \emph{chiral} (4,0) supersymmetry labelled by single-column Young tableaux. The proposed expressions originate from the sphere partition function of 3D circular quivers that include Wilson loops in antisymmetric representations for each node. Technically, we first use exact deconstruction to lift a 1D $\NN=2$ Fermi multiplet to a 2D $(2,0)$ Fermi multiplet; when integrated out, the first is known to be related to Wilson loop-operators in 3D \cite{Gomis:2006sb}, while the second to chiral surface defects in 4D \cite{Buchbinder:2007ar}. Then we use 3D information to propose a prescription that picks out contributions in the four-dimensional result labelled by single-column Young tableaux. These contributions are candidates for the partition function of corresponding 4D surface operators. To the best of our knowledge the 4D partition function of such defects has not been considered previously in the literature.

\item We obtain expressions for 4D $\NN = 2$ superconformal indices in the presence of codimension-one defects. We achieve this by starting from sphere partition functions of 3D circular quivers that include localised insertions of 3D matter/gauge fields. The implementation of the exact deconstruction process to such quiver defects is very simple and leads to 4D indices coupled to $S^3$ partition functions. From a 1D quantum-mechanics viewpoint the deconstructed defects are local supersymmetric operators on a circle.

\end{itemize}

The rest of this paper is organised as follows: In Sec.~\ref{exactdec} we provide a summary of the original dimensional-deconstruction proposal and a detailed description of both the technical as well as some conceptual points regarding exact deconstruction. We also review  how one can brane engineer our three-dimensional starting points and how to view deconstruction in that language. In Sec.~\ref{pf} we perform a warmup calculation, where we recover the index of 4D $\NN=2$ super-Yang--Mills (SYM) and super-QCD (SQCD) theories from the squashed-$S^3$ partition function of a 3D $\NN=2$ circular quiver. This is then extended in Sec.~\ref{vortex} to include vortex loops that lead (after deconstruction) to codimension-two defects with 2D $(2,2)$ supersymmetry. In this discussion, 1D and 2D defects are respectively coupled to 3D and 4D bulk theories before and after deconstruction. The extension of our procedure to 3D Wilson loops, and their potential interpretation in terms of codimension-two defects in 4D post deconstruction, appears in Sec.~\ref{wilson}. In Sec.~\ref{lines} we proceed to discuss how localised insertions of defects in the 3D quiver deconstruct codimension-one defects in one dimension higher. We finally conclude in Sec.~\ref{conclusions} with a summary of the main results and a list of possible future directions.

\section{Exact Deconstruction}\label{exactdec}

We begin with a general overview of dimensional deconstruction and its application at the level of supersymmetric sphere partition functions. We refer to the latter application as ``exact deconstruction'', since the partition functions include contributions from all orders in the coupling, and also to distinguish it from the original proposal that relates theories on flat backgrounds at different energy scales. Some of the details will be omitted at this stage, but will appear when we specialise to the cases of interest in the coming sections.

\subsection{Elements of Dimensional Deconstruction}

The dimensional-deconstruction prescription, in the absence of defects, is well known and can be summarised as follows:\footnote{Here we focus on supersymmetric QFTs, which are the main cases of interest in this paper. Also, in view of the applications that follow we consider unitary gauge groups. Clearly, these assumptions are not necessary features of deconstruction; e.g. one could use special unitary groups instead with no need to modify our prescription.} starting from a supersymmetric $N$-noded circular-quiver theory, where the nodes denote $U(k)$ vector multiplets with the same bare gauge coupling $G$\footnote{This is known as the ``orbifold point'' in the space of couplings.} and the links supersymmetric matter, one takes the theory onto the Higgs branch by allowing the scalars in the matter multiplets to simultaneously develop a vacuum expectation value (vev), $v {\one_{k\times k}}$. This has the effect of breaking the gauge group to a diagonal subgroup, $U(k)^N\to U(k)$. By flowing to low energies the degrees of freedom get reorganised into the KK modes of a theory in one dimension higher, compactified on a discretised circle with 
\begin{align}
  \label{eq:2}
  g^2_{\rm dec}\equiv\frac{G}{v}\to {\rm fixed}\;,\qquad 2 \pi \widehat{R} \equiv \frac{N}{G v}\to {\rm fixed}\;,\qquad a \equiv \frac{1}{G v}\;.
\end{align}
In these formulae, $\widehat{R}$ is the radius of the emerging circle, $a$ the lattice spacing and $g_{\rm dec}$ the emerging bare coupling \cite{ArkaniHamed:2001ca,Hill:2000mu}. Although the original deconstruction proposal relates four-dimensional quivers to five-dimensional theories on a circle, it can also be applied to lower-dimensional quivers \cite{Kaplan:2002wv}. It can be further generalised to theories of higher codimensionality by producing products of circles when starting from higher-dimensional periodic lattices in theory space \cite{ArkaniHamed:2001ie,Kaplan:2002wv}.

When the Higgs branch of the theory is not lifted by quantum corrections, e.g.\ when the quiver is superconformal or by adding an appropriate superpotential term, one can consider taking the combined limit
\begin{align}
  \label{eq:1}
  v \to \infty\;,\qquad N\to\infty\;,\qquad G\to \infty\;,
\end{align}
which sends $a\to 0 $ and recovers the continuum theory, while keeping $g_{\rm dec}$ and $\widehat{R}$ fixed; such examples may also exhibit supersymmetry enhancement for the deconstructed theory \cite{ArkaniHamed:2001ie,Csaki:2001em}.

\subsection{Deconstruction of Exact Partition Functions}

A proposal for implementing the  principle of dimensional deconstruction in the supersymmetric partition function of the $\mathcal N=2$ superconformal circular quiver on the ellipsoid, $S^4_{\epsilon_1, \epsilon_2}$, was given in \cite{Hayling:2017cva}. This partition function can be calculated using supersymmetric localisation \cite{Pestun:2007rz,Alday:2009aq,Pestun:2016zxk} or the refined topological-vertex formalism \cite{Iqbal:2007ii} and the result is schematically written as \begin{align}
  \label{eq:3}
\mathcal Z^{\rm quiver}_{S^4_{\epsilon_1, \epsilon_2}} = \prod_\alpha\int  [\mathrm d \lambda^{(\alpha)}] |Z_{\mathrm{4D}}^{(\alpha)}(\tau^{(\alpha)}, \lambda^{(\alpha)}, m^{(\alpha)}; \epsilon_1, \epsilon_2)|^2\;, 
\end{align}
where $Z^{(\alpha)}_{\mathrm{4D}} $ is a known function of the complexified gauge coupling $\tau^{(\alpha)}$, the squashing parameters $\epsilon_1, \epsilon_2$, the Coulomb-branch parameters $\lambda^{(\alpha)}$ and possibly some mass-deformation parameters $m^{(\alpha)}$, with $\alpha = \lfloor -\frac{N}{2}\rfloor +1, \ldots,\lfloor \frac{N}{2}\rfloor$ labelling the contribution from each quiver node \cite{Hayling:2017cva}. By identifying the parameters $\lambda^{(\alpha)}\to \lambda$, $m^{(\alpha)}\to m$, $\tau^{(\alpha)}\to \tau$ and assigning an additional mass parameter $m_\alpha=i\frac{\alpha}{\widehat{R}}$ to the $\alpha$-th node contribution---which accounts for the reorganisation of the massless spectrum into the massive KK tower obtained post deconstruction---the partition function of the six-dimensional (2,0) theory on $S_{\epsilon_1,\epsilon_2}^4\times T^2$ was obtained, as expected from \cite{ArkaniHamed:2001ie}.\footnote{See also \cite{Hayling:2018fmv}, where part of the partition function of the superconformal $\mathcal N=1$ toroidal-quiver theory was related to the partition function of Little String Theory on $S_{\epsilon_1,\epsilon_2}^4\times T^2$, again as expected from \cite{ArkaniHamed:2001ie}.}

However, the principle of dimensional deconstruction can be extended to exact partition functions with no specific reference to the number of spacetime dimensions. Assuming Lagrangian circular-quiver theories with the appropriate amount of supersymmetry,\footnote{A certain amount of supersymmetry is required to be able to use the localisation method on certain backgrounds such as $S^d$.} the $S^D$ partition functions can be put together in a modular fashion by accounting for the individual contributions coming from the various supersymmetric multiplets appearing in the quiver diagram. Each node will be labelled by an index $\alpha$ and the (integrand of the) corresponding partition function can generically depend on a coupling, Coulomb-branch parameter, mass-deformation parameter etc. Then, exact deconstruction can be implemented directly at the level of partition functions by:
\begin{itemize}
  \item[(1)] Identifying all parameters between the nodes (couplings, Coulomb-branch parameters, mass deformations etc.)---this reflects the breaking of the gauge symmetry, e.g.\ $U(k)^N\to U(k)$.
  \item[(2)] Including a mass parameter $m_\alpha = i\frac{\alpha}{\widehat{R}}$ for the $\alpha$-th node contribution---this captures the reorganisation of the quiver degrees of freedom into the KK modes of the deconstructed theory---and taking $N\to\infty$.
    
\end{itemize}

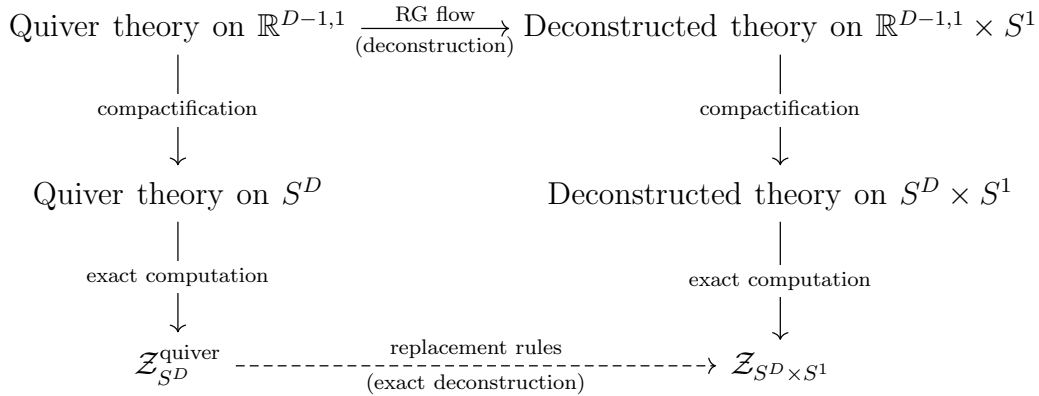
\begin{figure}[t]
\begin{center}
\begin{tikzcd}[column sep=huge, row sep=huge]
\textrm{Quiver theory on $\mathbb R^{D-1,1}$}\arrow[d, swap, "\text{compactification}" description]  \arrow{r}[above]{\text{RG flow}}[below]{\text{(deconstruction)}}  & \textrm{Deconstructed theory on $\mathbb R^{D-1,1}\times S^1$} \arrow[d, "\text{compactification}" description]\\
\textrm{Quiver theory on $S^D$} \arrow[d, swap, "\text{exact computation}" description]  & \textrm{Deconstructed theory on $S^{D}\times S^1$} \arrow[d, "\text{exact computation}" description]\\
\mathcal{Z}_{S^D}^{\mathrm{quiver}}\arrow[dashed]{r}[above]{\text{replacement rules}}[below]{\text{(exact deconstruction)}} & \mathcal{Z}_{S^D\times S^1} 
\end{tikzcd}
\end{center}
\caption{The deconstruction procedure at the level of exact partition
  functions. In this setup only one dimension is being deconstructed.}\label{dectable}
\end{figure}

Note that, contrary to the usual dimensional deconstruction, which describes a physical process relating two theories at different energy scales via a version of the Higgs mechanism, exact deconstruction should not be understood as a limiting procedure for theories defined on curved manifolds. This cannot be the case since the compactified version of the quiver theory does not have a moduli space of vacua, including vacua on the Higgs branch. Generic moduli are lifted by the couplings to the curvature of the (deformed) sphere. Instead, exact deconstruction should be viewed as a set of formal replacement rules for supersymmetric partition functions, accompanied by a set of mathematical identities that implement the dimensional lift. These replacement rules reflect the change in the spectrum after deconstructing the flat-space theory. We illustrate this set of relations in Fig.~\ref{dectable}.

In the present work we will demonstrate how this prescription can be applied to recover four-dimensional partition functions on $S_b^3 \times S^1_{\widehat{R}}$ from three-dimensional circular-quivers on the squashed three-sphere, $S_b^3$, as well as two-dimensional elliptic genera from the partition function of supersymmetric quantum mechanics on a circle; the latter encode the contributions of codimension-two defects. 

Let us close this discussion with a comment that has to do with the legitimacy of the continuum deconstruction limit \eqref{eq:1}. Unlike the circular-quiver theories of \cite{ArkaniHamed:2001ie,Hayling:2017cva,Hayling:2018fmv}, the 3D quivers of this paper are not superconformal, so one needs to make sure that quantum effects do not lift the Higgs branch---the existence of which is necessary for deconstruction---at low energies. Single-noded $\NN = 2$ SQCD theories in three dimensions  with $N_f \geq N_c$ fundamental/anti-fundamental pairs are believed to have distinct Higgs and Coulomb branches, even after incorporating quantum corrections. Moreover, at the intersection of these branches such theories flow to interacting critical points in the IR \cite{deBoer:1997kr,Aharony:1997bx}. In all the examples that we will consider in this paper, we have circular quivers where $N_f\geq N_c$ is obeyed at each node. We expect that the above statements about the non-lifting of the Higgs and Coulomb branches extend to the full theory. Although we will not present rigorous arguments to this effect, we will see in the upcoming sections that this picture leads to sensible results. Related statements about the low energy dynamics of 3D $\NN=4$ quivers can be found in \cite{Gaiotto:2008ak}. Further evidence that supports the validity of dimensional deconstruction in the cases that we consider is provided in the next subsection with a suitable embedding in string theory.

\subsection{Brane Engineering}

\begin{table}[t]
\centering
\begin{tabular}{|c|c|c|c|c|c|c|c|c|c|c|c|}
\hline
&$x^0$ & $x^1$ &$x^2$ &$x^3$ &$x^4$ &$x^5$ &$x^6$ &$x^7$ &$x^8$ &$x^9$ \\
  \hline
  D3  &$-$&$-$&$-$&$-$&$\cdot$&$\cdot$&$\cdot$&$\cdot$&$\cdot$&$\cdot$\\
  \hline
  NS5  &$-$&$-$&$-$&$\cdot$&$\cdot$&$\cdot$&$\cdot$&$-$&$-$&$-$\\
\hline
 D5   &$-$&$-$&$-$&$\cdot$&$-$&$-$&$-$&$\cdot$&$\cdot$&$\cdot$\\
  \hline
 D1   &$-$&$\cdot$&$\cdot$&$\cdot$&$\cdot$&$\cdot$&$-$&$\cdot$&$\cdot$&$\cdot$\\
\hline
  NS5$'$  &$-$&$\cdot$&$\cdot$&$\cdot$&$-$&$-$&$\cdot$&$-$&$-$&$-$\\
\hline
$A_{N-1}$  &$\cdot$&$\cdot$&$\cdot$&$\cdot$&$-$&$-$&$\cdot$&$\cdot$&$-$&$-$\\
\hline
\end{tabular}
\caption{Brane configuration in type IIB string theory engineering 3D vortex-loop operators. The $\mathbb Z_N$ orbifold acts in the directions $x^4,x^5,x^8,x^9$. The D3s are stretched between NS5s along an interval $L_3$ in the $x^3$ direction. The D1s are stretched between the D3s and NS5s and/or NS5$'$s along the $x^6$ direction. }
\label{IIB}
\end{table}

Before implementing the technical steps of exact deconstruction it is useful to have complementary evidence that dimensional deconstruction works unobstructed in the full physical theory. In many cases such evidence is encoded naturally in string-theory embeddings. All 3D theories that we will be interested in in this paper arise in the low-energy limit of type IIB string theory configurations that involve branes of the type listed in Tables \ref{IIB} and \ref{IIBW}. The precise combinations of these ingredients depend on the specific example under study and we will spell out the pertinent details in the upcoming sections as needed. In the rest of this subsection we summarise the key features of these constructions.

The low-energy dynamics of multiple D3s suspended between NS5s along an interval $L_3$ in the $x^3$ direction are captured by a three-dimensional gauge theory. The inclusion of D5 branes that intersect the D3s corresponds to the introduction of flavour. In order to engineer vortex loops, which are related to codimension-two defects with \emph{non-chiral} 2D supersymmetry through dimensional reduction, one needs to introduce D1s that stretch between the D3s and an NS5 and/or NS5$'$s along the $x^6$ direction, oriented as in Tab.~\ref{IIB}. In order to engineer Wilson loops, which are related to codimension-two defects with \emph{chiral} 2D supersymmetry through dimensional reduction, one needs to introduce F1s that stretch between the D3s and a D5 and/or D5$'$ along the $x^7$ direction, oriented as in Tab.~\ref{IIBW}. In this fashion, one can engineer a wide variety of examples with $\mathcal N=4$ supersymmetry; see e.g.\ \cite{Assel:2015oxa}. Placing various combinations of the above ingredients on a $\mathbb C^2/\mathbb Z_N$ orbifold singularity, $A_{N-1}$, leads to a circular-quiver gauge theory in three dimensions with $\mathcal N=2$ supersymmetry and vortex/Wilson loops at low energies \cite{Douglas:1996sw,Giveon:1998sr}.

In this setup deconstruction is simply realised by taking the combined brane system off the $A_{N-1}$ singularity and into the orbifolded space, which is locally $\mathbb R^3 \times S^1_{\widetilde{R}}$, by a distance $d$. In the presence of flavour/defect branes this is carried out in one of the directions $x^8$ or $x^9$, with $x^9$ or $x^8$  respectively compactified on a circle of radius $\widetilde{R}$. The continuum limit \eqref{eq:1} involves taking the string length scale $l_s \to 0$ and $N\to \infty$ while keeping the string coupling $g_s \to\mathrm{fixed}$ and $\widetilde{R}/l_s^2 \equiv d/N l_s^2 \to\mathrm{fixed}$. An equivalent description is in terms of a T-dual system with the various D-branes wrapping a fixed-sized circle $\widehat{R} \equiv l_s^2/\widetilde{R}$ and string coupling $g_s' = g_s \widehat{R}/l_s$ \cite{ArkaniHamed:2001ie,Hayling:2017cva}. In order to keep the 3D gauge coupling $1/g_{\mathrm{3D}}^2 \equiv L_3 l_s/g_s$ tuneable, as required by deconstruction, one needs to take $L_3 \to \infty$ as $l_s \to 0$. Then, the 4D gauge coupling $1/g_{\mathrm{4D}}^2 \equiv L_3/g_s' l_s = L_3/g_s \widehat{R} \to\infty$ in the limit, resulting in a weakly-coupled 4D gauge theory (with matter and/or defects).

\section{4D Indices from $S^3$ Partition Functions}\label{pf}

\begin{table}[t]
\centering
\begin{tabular}{|c|c|c|c|c|c|c|c|c|c|c|c|}
\hline
&$x^0$ & $x^1$ &$x^2$ &$x^3$ &$x^4$ &$x^5$ &$x^6$ &$x^7$ &$x^8$ &$x^9$ \\
  \hline
  D3  &$-$&$-$&$-$&$-$&$\cdot$&$\cdot$&$\cdot$&$\cdot$&$\cdot$&$\cdot$\\
\hline
  NS5  &$-$&$-$&$-$&$\cdot$&$\cdot$&$\cdot$&$\cdot$&$-$&$-$&$-$\\
\hline
 D5   &$-$&$-$&$-$&$\cdot$&$-$&$-$&$-$&$\cdot$&$\cdot$&$\cdot$\\
  \hline
 F1   &$-$&$\cdot$&$\cdot$&$\cdot$&$\cdot$&$\cdot$&$\cdot$&$-$&$\cdot$&$\cdot$\\
  \hline
  D5$'$  &$-$&$\cdot$&$\cdot$&$\cdot$&$-$&$-$&$-$&$\cdot$&$-$&$-$\\
  \hline
  $A_{N-1}$  &$\cdot$&$\cdot$&$\cdot$&$\cdot$&$-$&$-$&$\cdot$&$\cdot$&$-$&$-$\\
  \hline
\end{tabular}
\caption{Brane configuration in type IIB string theory engineering 3D Wilson-loop operators. The $\mathbb Z_N$ orbifold acts in the directions $x^4,x^5,x^8,x^9$. The D3s are stretched between NS5s along an interval $L_3$ in the $x^3$ direction. The F1s are stretched between the D3s and D5s and/or D5$'$s  along the $x^7$ direction.}
\label{IIBW}
\end{table}

The exact-deconstruction procedure can be straightforwardly applied in the context of 3D quiver-gauge theories and their partition functions on the $U(1)\times U(1)$ isometric hyper-ellipsoid, $S^3_b$. The latter is given by the equation
\begin{align}
  \label{eq:4}
\tilde \ell^{-2} (X_1^2 +X_2^2) +\ell^{-2} (X_3^2 + X_4^2 ) = 1 \;,
\end{align}
with the squashing parameter defined typically as the dimensionless ratio $b=\sqrt{\tilde \ell \over \ell}$; the round three sphere is then recovered in the limit $b\to 1$. In what follows, we will make repeated use of two purely-imaginary numbers, $\omega_1$ and $\omega_2$, defined as $\omega_1 = i b$, $\omega_2=i b^{-1}$. Supersymmetric partition functions on the ellipsoid are readily calculated using supersymmetric localisation \cite{Kapustin:2009kz,Hama:2010av,Hama:2011ea}. The answer can be neatly organised by observing that it factorises into individual contributions from vector and chiral multiplets---in the appropriate representations of the Lie algebra---over which one needs to perform a final matrix integral for any symmetries that are gauged. We summarise our notation and the key ingredients in the following subsection.

\subsection{Building Blocks of Sphere Partition Functions}\label{3Dblocks}

The partition function of 3D $\mathcal{N}=2$ theories on  $S^3_b$ can be constructed using the vector- and chiral-multiplet partition functions
\begin{align}
\begin{split}
\label{3DN2cont}
Z^{\mathcal{N}=2}_{\rm{vec}}(\lambda) =& \prod_{\beta \in \rm{Adj}} \widehat{\Gamma}_h\left(\langle \beta, \lambda \rangle \big| \omega_1, \omega_2\right)^{-1}\cr
=&\,\widehat{\Gamma}_h\left(0|\omega_1, \omega_2\right)^{-\rm{rank}(\mathfrak g)}\prod_{\beta \in \Delta} \widehat{\Gamma}_h\left( \langle \beta, \lambda\rangle \big| \omega_1, \omega_2\right)^{-1}\;, \cr 
Z^{\mathcal{N}=2}_{\rm{chi}}(\lambda,r) =& \prod_{\beta \in \mathcal{R}} \Gamma_h\left(r\omega_+ - \langle \beta, \lambda\rangle \big| \omega_1, \omega_2\right) \;,
\end{split}
\end{align}
where $\Gamma_h$ and $\widehat\Gamma_h$ are hyperbolic Gamma functions---defined in App.~\ref{math}. Here  $\mathcal{R}$ is a generic representation in any product of gauge or global symmetries and $\lambda$ is associated with a set of chemical potentials corresponding to these symmetries, $\beta$ takes values in the weights of the representation $\mathcal R$\footnote{For the adjoint representation the product over $\beta$ can be reduced to a product over the roots $\Delta$ and Cartans $\mathfrak h$.} and $r$ denotes the chiral-multiplet $U(1)_R$ charge. The product $\langle\beta , \lambda \rangle$ is an inner product in weight space. We are also using the combination $\omega_+ = \frac{\omega_1+ \omega_2}{2}$.

Note that we can combine the vector-multiplet contributions with the matrix-integral Haar measure, $\Delta^{\rm Haar}(\lambda) =  \prod_{\beta\in \Delta} i \langle \beta, \lambda\rangle $, to write
\begin{align}
\Delta^{\rm{Haar}}(\lambda)  Z^{\mathcal{N}=2}_{\rm{vec}}(\lambda) =\widehat\Gamma_h(0|\omega_1,\omega_2)^{-\rm{rank}(\mathfrak g)}\prod_{\beta \in \Delta} \Gamma_h\left( \langle \beta, \lambda \rangle \big| \omega_1, \omega_2\right)^{-1}\;.
\end{align}

The above ingredients can be used to construct 3D $\mathcal{N}=4$ multiplet contributions. For the vector multiplet we combine an $\mathcal{N}=2$ vector multiplet with an adjoint $\mathcal{N}=2$ chiral multiplet with $r = 1$. For an adjoint chiral $Z^{\mathcal{N}=2}_{\rm{Adj\;chi}}(\lambda, r = 1) = 1$,\footnote{This is due to the result enjoying a symmetry under root reflections.} hence 
\begin{align}
  \Delta^{\rm{Haar}}(\lambda)  Z^{\mathcal{N}=4}_{\rm{vec}}(\lambda) =\widehat\Gamma_h(0|\omega_1,\omega_2)^{-\rm{rank}(\mathfrak g)}\prod_{\beta \in \Delta} \Gamma_h\left( \langle \beta,  \lambda\rangle \big| \omega_1, \omega_2\right)^{-1}\;,
\end{align}
which is the same function as in $\mathcal{N}=2$. For the 3D $\mathcal{N}=4$ hypermultiplet, one simply takes two $\mathcal{N}=2$ chiral multiplets in conjugate representations, both with $r=\frac{1}{2}$. Thus
\begin{align}
  Z^{\mathcal{N}=4}_{\rm{hyp}}(\lambda) = \prod_{\beta \in \mathcal{R}} \Gamma_h\left(\frac{\omega_+}{2}  -  \langle \beta, \lambda \rangle \big| \omega_1, \omega_2\right)\Gamma_h\left(\frac{\omega_+}{2} +  \langle \beta, \lambda \rangle \big| \omega_1, \omega_2\right)\;.
\end{align}

Post deconstruction, and for the appropriate choice of matter content, the $S^3$ partition functions of 3D $\NN=2$ theories will lift to the superconformal index of 4D $\mathcal N=2$ theories \cite{Kinney:2005ej}. For superconformal $\NN=2$ theories in Euclidean $\mathbb R^4$, and in radial quantisation, this index is defined as the trace (we use the conventions of \cite{Gadde:2013ftv})
\begin{align}
  \label{eq:8}
  \mathcal I_{\mathrm{4D}} = \Tr (-1)^Fe^{- \beta (E - h_{01}-h_{23}-2R -r)}\fp^{h_{23}-r}\fq^{h_{01}-r}\ft^{R+r}\;,
\end{align}
where $F$ is the fermion number, $E$ the conformal dimension, $h_{01} = j_1 + j_2$ and  $h_{23} = -j_1 + j_2$ are rotation generators along the two planes $01$ and $23$ of $\mathbb R^4$, $R$ is the $SU(2)_R$ Cartan, while $r$ is the $U(1)_r$ Cartan. One can also make the change of variables $\ft\to \mathfrak v \sqrt{\fp \fq}$ to arrive at
\begin{align}
  \label{eq:9}
  \mathcal I_{\mathrm{4D}} = \Tr (-1)^Fe^{- \beta (E - h_{01}-h_{23}-2R -r)}\fp^{h_{23}+\frac{1}{2}(R -r)}\fq^{h_{01}+\frac{1}{2}(R-r)}\mathfrak v^{R+r}\;,
\end{align}
which in the limit $\mathfrak v\to 1 $ reduces to an $\mathcal N=1$ superconformal index \cite{Dolan:2008qi}.\footnote{Via the operator-state map the superconformal index also admits a presentation in terms of a twisted partition function on $S^3 \times S^1$ with supersymmetric boundary conditions and  Hamiltonian $H = E - h_{01}-h_{23}-2R -r$ \cite{Romelsberger:2005eg}. This definition is more general as it is also applicable to non-conformal theories, for which the name superconformal index is not quite appropriate. However, since this quantity is independent of the coupling and the theory eventually flows to some SCFT at an IR fixed point, we will still use this nomenclature through a mild abuse of language.}

The superconformal index of Lagrangian theories can itself be neatly organised in terms of separate contributions from vector multiplets and chiral multiplets, all of which are again expressible via special functions. The special function that dominates the 4D superconformal index is the elliptic Gamma function $\Gamma_e(z| {\mathfrak p}, {\mathfrak q})$ (also defined in App.~\ref{math}). In what follows we explain how deconstruction recovers all the details of the 4D superconformal index from three-dimensional data.

\subsection{Deconstruction of 4D $\mathcal N=2$ Pure SYM Theory}\label{N2decsec}

\begin{figure}[t]
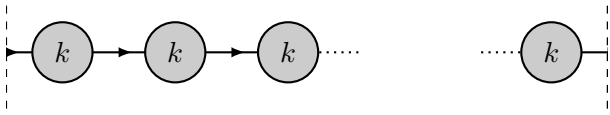
 
\begin{center} 
  \drawquiverpuresym
\end{center}
\caption{3D $\mathcal N=2$ circular quiver with $N$ nodes. The nodes denote $U(k)$ vector multiplets, while the links bifundamental-chiral multiplets with assigned R charge $r=1$. The endpoints are to be understood as periodically identified.}\label{fig:2} 
\end{figure}

Let us begin with the simplest possible example, which will allow us to highlight the main points of the exact-deconstruction procedure. In the brane-system description of Tab.~\ref{IIB} this involves $kN$ D3s on the $A_{N-1}$ singularity, suspended between two NS5s that are separated by an interval of size $L_3$. In the string-decoupling limit this setup gives rise to the three-dimensional quiver-gauge theory with $\mathcal N=2$ supersymmetry illustrated in Fig.~\ref{fig:2}. Up to factors of $\omega_1, \omega_2$, the partition function of this theory on the ellipsoid, $S^3_b$, can be immediately written down with the help of \eqref{3DN2cont} as
\begin{align}\label{simplequiver}
\mathcal Z^{\textrm{quiver}}_{\mathrm{3D}} &=\prod_{\alpha}\frac{1}{k!}\int \prod_{b=1}^{k} \mathrm d \sigma_b^{(\alpha)} \Delta^{\text{Haar}} \left(\sigma^{(\alpha)}\right) \prod_{b,c=1}^{k} \frac{\Gamma_h\left(\omega_+ + \sigma^{(\alpha)}_b - \sigma^{(\alpha+1)}_c \Big| \omega_1, \omega_2\right)}{ \widehat{\Gamma}_h\left(\sigma^{(\alpha)}_b - \sigma^{(\alpha)}_c \Big| \omega_1, \omega_2\right) }\;,
\end{align}
where the $\sigma^{(\alpha)}$ are (dimensionless versions of) Coulomb-branch parameters for the gauge group at each node.\footnote{The dimensions of the Coulomb-branch parameters can be restored by multiplying with  $(\ell\tilde \ell)^{-\frac{1}{2}}$, where $\ell,\tilde \ell$ are the length scales that appear in the definition of the ellipsoid \eqref{eq:4}.} 

As we have already described, the exact-deconstruction prescription has two key elements:
\begin{itemize}
\item[(1)] We should identify all $\sigma^{(\alpha)}\to \sigma$. This encodes the breaking of the gauge symmetry $U(k)^N\to U(k)$.
\item[(2)] We should shift all the arguments of the hyperbolic Gamma functions by $\Gamma_h(x)\to \Gamma_h(x + \frac{\alpha}{R})$ and take $N\to\infty$. This encodes the reorganisation of the spectrum into the KK modes of the higher-dimensional theory with mass parameters $m_\alpha =i \frac{\alpha}{R}$. In these formulae, $R$ is a dimensionless parameter, which is related to the dimensional radius $\widehat R$ of the deconstructed $S^1$, via $R=\frac{2\pi\widehat R}{\sqrt{\ell \tilde \ell}}$.
\end{itemize}
Implementing these steps leads to the partition function
\begin{align}
  \label{eq:6}
\mathcal Z^{\textrm{Dec}}_{\mathrm{3D}} &= \frac{1}{k!}\int \prod_{b=1}^{k}\mathrm d \sigma_b \prod_{\alpha}\frac{\Gamma_h\left(\omega_++ \frac{\alpha}{R}\Big |\omega_1, \omega_2\right )^{k}}{\widehat\Gamma_h(\frac{\alpha}{R}|\omega_1,\omega_2)^{k}} \prod_{b\neq c} \frac{\Gamma_h\left(\omega_+  + \sigma_b - \sigma_c  + \frac{\alpha}{R}\Big| \omega_1, \omega_2\right)}{ \Gamma_h\left(\sigma_b - \sigma_c + \frac{\alpha}{R}\Big| \omega_1, \omega_2\right) }\;. 
\end{align}

\begin{figure}[t]
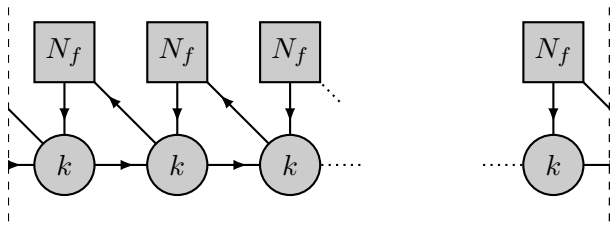
 
\begin{center} 
\drawquiversqcd
\end{center}
\caption{3D $\mathcal N=2$ circular quiver with $N$ gauge and flavour nodes. The circular nodes denote $U(k)$ vector multiplets, while the squares $U(N_f)$ flavour groups. The links between circles encode bifundamental-chiral multiplets with assigned R charge $r=1$, while the ones between circles and squares (anti)fundamental chirals with assigned R charge $r=\frac{1}{2}$. The endpoints are to be understood as periodically identified.}\label{fig:5} 
\end{figure}

At this stage one observes the following identity between hyperbolic and elliptic Gamma functions
\beq
\label{mathagmain}
\prod_{\alpha = -\infty}^\infty \Gamma_h\left(x + \frac{\alpha}{R}\bigg| \omega_1 , \omega_2\right) = \mathfrak x^2 \left(\fp \fq\right)^{-\frac{1}{2}} \Gamma_e (\mathfrak x |\fp, \fq)
\eeq
for $\mathfrak x=e^{2\pi i R x}$, $\fp = e^{2\pi i R \omega_1}$, $\fq= e^{2\pi i R \omega_2}$.\footnote{Further comments about this identity appear in App.\ \ref{math}. This identity also extends to the hatted versions of the hyperbolic and elliptic gamma functions.} The proof of this formula follows from a straightforward manipulation of the infinite-product representation of the hyperbolic Gamma functions. Up to dimensionless combinations involving factors of $\omega_1, \omega_2$ and $R$,\footnote{We will ignore such factors from now on.} one can therefore write
\begin{align}
  \label{eq:5}
  \mathcal Z^{\textrm{Dec}}_{\mathrm{3D}} &=\frac{1}{k!}\oint \prod_{b=1}^{k} \frac{\mathrm d v_b}{2 \pi i v_b}\frac{\Gamma_e(\sqrt{\fp \fq}| \fp , \fq)^{k}}{\widehat\Gamma_e(1|\fp,\fq)^{k} } \prod_{b\neq c} \frac{\Gamma_e\left(\sqrt{\fp \fq}v_b v_c^{-1}\Big| \fp,\fq\right)}{\Gamma_e\left(v_bv_c^{-1}\Big|\fp,\fq\right) }\;,  
\end{align}
where $v_b = e^{2 \pi i R \sigma_b}$. Finally, using $\widehat\Gamma_e(1 |\fp,\fq) = (\fp;\fp)^{-1} (\fq;\fq)^{-1}$ and recognising that $\Gamma_e(\sqrt{\fp \fq}| \fp , \fq) = 1$, as well as $\prod_{b\neq c}\Gamma_e(\sqrt{\fp \fq} v_b v_c^{-1}| \fp , \fq) = 1$, we arrive at
\begin{align}
  \label{eq:7}
\mathcal Z^{\textrm{Dec}}_{\mathrm{3D}} &=  \frac{1}{k!} (\fp;\fp)^{k} (\fq;\fq)^{k}\oint \prod_{b=1}^{k}\frac{\mathrm d v_b}{2 \pi i v_b} \prod_{b\neq c}\Gamma_e\left(v_bv_c^{-1}\Big|\fp,\fq\right)^{-1}\;,
\end{align}
which reproduces precisely the expression for the $\mathcal N=1$ superconformal index of a 4D $\mathcal N=2$ vector multiplet.\footnote{This corresponds to taking the $\mathfrak{t}\to\sqrt{\mathfrak{p} \mathfrak{q} } $ limit in the expressions of \cite{Gadde:2013ftv}.} Note that, although the supersymmetry of the deconstructed theory is double that of the original one, the 3D partition function has a single R-charge chemical potential and can therefore only provide the $\mathfrak v\to 1$ limit of the full $\mathcal N=2$ superconformal index \eqref{eq:9}.\footnote{It is possible to deconstruct the additional R-symmetry fugacity present in the $\mathcal N=2$ index by turning on appropriate real-mass terms in the $S^3$ partition function. Real masses in the $S^3$ partition function are related to background-$U(1)$ gauge fields in the $S^3 \times S^1$ partition function through dimensional reduction  \cite{Festuccia:2011ws}. The $S^3 \times S^1$ partition function in turn provides an alternative presentation of the index via the state-operator map, with generic background-$U(1)$ gauge fields mapping to global fugacities. Therefore, real masses in 3D will generically deconstruct global fugacities in 4D. In particular, turning on real masses, where the accompanying charges specifically correspond to the $R+r$ combination that appears in the $\mathcal N=2$ index (-1 for  bifundamental chiral multiplets and $\frac{1}{2}$ for the (anti)fundamental chiral multiplets of the next example), reproduces the contributions of the fugacity $\mathfrak v$.}

\subsection{Deconstruction of 4D $\mathcal N=2$ SQCD}

One can further enrich the previous example by adding flavour. In the brane picture this corresponds to adding $N_f N$ D5 branes to the $kN$ D3s suspended between two NS5s in the presence of the $A_{N-1}$ orbifold singularity, along the directions listed in Tab.~\ref{IIB}. In the string-decoupling limit this gives rise to the quiver of Fig.~\ref{fig:5}, which preserves $\mathcal N=2$ supersymmetry in three dimensions. The $S^3_b$ partition function of this circular-quiver theory is given by the expression
\begin{align}
\begin{split}
\mathcal Z^{\textrm{quiver}}_{\mathrm{3D}} &=\prod_{\alpha}\frac{1}{k!}\oint \prod_{b=1}^{k} \mathrm d \sigma_b^{(\alpha)} \Delta^{\text{Haar}} \left(\sigma^{(\alpha)}\right) \prod_{b,c=1}^{k} \frac{\Gamma_h\left(\omega_+ + \sigma^{(\alpha)}_b - \sigma^{(\alpha+1)}_c \Big| \omega_1, \omega_2\right)}{ \widehat{\Gamma}_h\left(\sigma^{(\alpha)}_b - \sigma^{(\alpha)}_c \Big| \omega_1, \omega_2\right) }\\
                                                        &\times \prod_{b=1 }^k\prod_{j=1}^{N_f} \Gamma_h\left(\frac{1}{2}\omega_+ -  \mu^{(\alpha)}_j+ \sigma^{(\alpha)}_b \Big| \omega_1, \omega_2\right) \Gamma_h\left(\frac{1}{2}\omega_+  - \sigma^{(\alpha+1)}_b +  \mu_j^{(\alpha)} \Big| \omega_1, \omega_2\right) \;.
\end{split}
\end{align}

In order to perform the dimensional-deconstruction procedure, even in the flat-space case, one first needs to explicitly break the flavour group to its diagonal subgroup $U(N_f)^N\to U(N_f)$ \cite{Csaki:2001em}; this has the effect of identifying all the square nodes in Fig.~\ref{fig:5} and sending $\mu_j^{(\alpha)}\to\mu_j$.\footnote{From the point of view of the D3-brane theory the modes describing the positions of the D5s are non-dynamical (they are ``frozen out'') and should be viewed as parameters of the low-energy theory. The identification of the square nodes in Fig.\ \ref{fig:5} corresponds to taking the D5s off the orbifold singularity before implementing the dynamical Higgsing through moving the D3s.} 
Apart from this additional detail, the exact-deconstruction procedure can be implemented as in the previous example, by identifying all the Coulomb-branch parameters and shifting all the arguments of the hyperbolic-gamma functions. This leads to the formula
\begin{align}
\begin{split}
\mathcal Z^{\textrm{Dec}}_{\mathrm{3D}} &= \frac{1}{k!}\int \prod_{b=1}^{k}\mathrm d \sigma_b \prod_{\alpha}\frac{\Gamma_h(\omega_++ \frac{\alpha}{R}\Big |\omega_1, \omega_2)^{k}}{\widehat\Gamma_h(\frac{\alpha}{R}|\omega_1,\omega_2)^{k}} \prod_{b\neq c} \frac{\Gamma_h\left(\omega_+  + \sigma_b - \sigma_c  + \frac{\alpha}{R}\Big| \omega_1, \omega_2\right)}{ \Gamma_h\left(\sigma_b - \sigma_c + \frac{\alpha}{R}\Big| \omega_1, \omega_2\right) }
\\&\times \prod_{b=1 }^k\prod_{j=1}^{N_f} \Gamma_h\left(\frac{1}{2}\omega_+ \mp \sigma_b \pm  \mu_j + \frac{\alpha}{R}\Big| \omega_1, \omega_2\right) 
\;.
\end{split}
\end{align}

Finally, using \eqref{mathagmain} we arrive at
\begin{align}\label{final}
  \mathcal Z^{\textrm{Dec}}_{\mathrm{3D}} &=  \frac{1}{k!} (\fp;\fp)^{k} (\fq;\fq)^{k}\oint \prod_{b=1}^{k}\frac{\mathrm d v_b}{2 \pi i v_b} \prod_{b\neq c}\Gamma_e\left(v_bv_c^{-1}\Big|\fp,\fq\right)^{-1}\prod_{b=1 }^k\prod_{j=1}^{N_f} \Gamma_e\left((\fp\fq)^{\frac{1}{4}} (v_b s_j^{-1})^{\pm} \Big| \fp,\fq\right) \;,
\end{align}
where $s_j = e^{2 \pi i R \mu_j}$.\footnote{In the above equation and
  hereafter we employ the commonplace notation $\Gamma_e(\mathfrak
  x^{\pm})\equiv \Gamma_e(\mathfrak x)\Gamma_e(\mathfrak x ^{-1})$.}
When $N_f = 2k$, our $U(k)$ theory is not conformally
invariant. However, it flows to 4D $\mathcal N = 2$ SCQCD with a weak $U(1)$ gauging times a free abelian vector multiplet in the IR. The IR theory admits an honest $\mathcal N=1$ superconformal index and, upon a reparametrisation $\tilde u_b = u_b a^{-\frac{1}{k}}$,\footnote{The $\tilde u_b$ are $k$ $SU(k)$ fugacities obeying $\prod_{b =1}^k \tilde u_b= 1$, while $a$ denotes the fugacity for the free $U(1)$ sector.} the result \eqref{final} is in explicit agreement with this quantity.

\subsection{Comments on Further Examples}

It is clear that with the ingredients that we have provided one can deconstruct a host of more complicated examples in 4D with $\mathcal N=2$ supersymmetry. Exact deconstruction operates in a modular fashion on each hyperbolic Gamma factor of the $S^3_b$ partition function to reconstruct the corresponding elliptic Gamma factor of the 4D superconformal index. 

In recent years several authors have examined dimensional reduction at the level of exact partition functions as a bridge between exact quantities for theories in different spacetime dimensions. The degeneration limits of 4D superconformal indices to 3D $S^3$ partition functions were discussed originally in a QFT context in \cite{Dolan:2011rp,Gadde:2011ia}. This study becomes particularly interesting when it is combined with non-perturbative dualities \cite{Niarchos:2012ah,Aharony:2013dha}.

The exact deconstruction of this section should be seen as a useful complementary relation between 3D and 4D physics. Compared to dimensional reduction, a potentially-interesting feature of deconstruction\footnote{Or exact deconstruction compared to superconformal-index degeneration at the level of partition functions.} that we would like to highlight in this context is the following. In some cases dimensional reduction of a 4D theory leads to a 3D theory with accidental symmetries at strong coupling, whose $S^3$ partition function---defined as an integral---is not well behaved. In such cases the 3D quiver-gauge theory that deconstructs the 4D theory does not have to be equally problematic. A prominent example can be found in $\NN=4$ SYM theory.   

The dimensional reduction of 4D $\NN=4$ SYM theory gives 3D $\NN=8$ SYM theory, a theory whose IR superconformal symmetry is not visible in the UV description. In contrast, the dimensional deconstruction of 4D $\NN=4$ SYM theory can be performed with a circular 3D $\NN=4$ quiver of the type depicted in Fig.\ \ref{fig:2}, where each circle denotes now a 3D $\NN=4$ vector multiplet and each link a 3D $\NN=4$ hypermultiplet (see also the 3D part of the upcoming Fig.\ \ref{3DN8Orb}). This is a balanced 3D quiver-gauge theory with a well-defined $S^3$ partition function. Exact deconstruction in this case recovers the superconformal index of 4D $\NN=4$ SYM along the lines of the previous discussion.

\section{4D Surface Defects from 3D Vortex Loops}\label{vortex}

We next examine how exact deconstruction works in the presence of codimension-two defects. Codimension-two defects in three-dimensional QFTs are line operators. In four-dimensional QFTs they are surface operators. We will explain how exact deconstruction lifts $S^3_b$ partition functions with supersymmetric line-operator insertions in 3D $\NN=2$ (or 3D $\NN=4$) quiver-gauge theories to $S^1 \times S_b^3$ partition functions with surface-operator insertions in 4D $\NN=2$ (or 4D $\NN=4$) gauge theories. Our line defects will always wrap the fibre of the (squashed) Hopf fibration $S^1_b\hookrightarrow S^3_b\rightarrow S^2_b$, and will be situated at the North or South pole of the $S^2_b$, as in \cite{Assel:2015oxa}. Note that the squashing deformation breaks the same supersymmetry as the line defect. So even though in flat space the line defects are $\frac{1}{2}$-BPS, on the $S^3_b$ they do not break any additional supersymmetry. 

In 3D $\NN=4$ theories one can consider two types of $\frac{1}{2}$-BPS line defects supported on a straight line in flat space. The first type preserves the 1D $\NN=\mathrm{4A}$ supersymmetry algebra, which arises from the dimensional reduction of 2D $\NN=(2,2)$ supersymmetry.\footnote{We note in passing that the dimensional reduction of 2D $\NN=(2,2)$ surface defects in 4D $\NN=2$ gauge theories to three dimensions has been discussed in \cite{Bullimore:2014nla}.} $\frac{1}{2}$-BPS vortex lines in 3D $\NN=4$ QFTs realise this symmetry. The second type preserves the 1D $\NN=\mathrm{4B}$ supersymmetry algebra, which arises from the dimensional reduction of 2D $\NN=(4,0)$ supersymmetry. $\frac{1}{2}$-BPS Wilson lines in 3D $\NN=4$ QFTs are examples of this type. A review of pertinent details can be found in \cite{Assel:2015oxa}. 

Defects of each of these types can generically be studied by coupling a 1D supersymmetric quantum mechanics theory (SQM), supported on the defect worldvolume, to the 3D bulk gauge theory of interest (here a quiver-gauge theory). The supersymmetry of the SQM defines the type of the defect. The typical coupling between the defect and the bulk proceeds via gauging 1D global symmetries with the vector multiplets of the bulk gauge theory. The coupling may also include superpotential terms involving defect and bulk matter fields.

Similarly, in 3D $\NN=2$ theories one can consider $\frac{1}{2}$-BPS line defects preserving either 1D $\NN=\mathrm{2A}$ supersymmetry (that arises from the dimensional reduction of 2D $\NN=(1,1)$ supersymmetry) or $\frac{1}{2}$-BPS defects preserving 1D $\NN=\mathrm{2B}$ supersymmetry (that arises from the dimensional reduction of 2D $\NN=(0,2)$ supersymmetry). 

In this paper we will consider line defects in 3D supersymmetric quiver-gauge theories that preserve the $\NN=\mathrm{4A}$, $\NN=\mathrm{4B}$ or $\NN=\mathrm{2B}$ supersymmetries. The $\NN=\mathrm{2A}$ supersymmetry will not play any r\^ole in the cases that we study. In the rest of this section we will discuss how to use exact deconstruction to lift the partition functions of vortex loops with 1D $\NN=\mathrm{4A}$ or $\NN=\mathrm{2B}$ supersymmetry in a 3D bulk, to indices for surface defects with 2D $\NN=(4,4)$ or 2D $\NN=(2,2)$ supersymmetry in a 4D bulk. Note that in this fashion we will bypass the discussion of dimensionally deconstructing the 4D-2D system from 3D-1D at the level of Lagrangians for the flat-space theories, comparing instead the theories  directly at the level of sphere partition functions. The case of Wilson loops is relegated to Sec.~\ref{wilson}.

\subsection{Building Blocks of Partition Functions with Vortex-Loop Insertions}\label{defingr}

In the presence of codimension-two defects the $S^3_b$ partition functions will have, in addition to the  3D-multiplet contributions that we discussed in Sec.~\ref{3Dblocks}, contributions coming from the 1D defect theory, which is supported on a circle of (dimensionless) radius $\omega^{-1}$;\footnote{In toric coordinates the metric on the ellipsoid is 
$$ds^2=\left( \tilde \ell^2 \cos^2\theta + \ell^2 \sin^2\theta \right) d\theta^2 + \tilde \ell^2 \sin^2\theta\, d\varphi_1^2+\ell^2 \cos^2\theta\, d\varphi_2^2~,$$
where $\theta \in [0,\frac{\pi}{2}]$ and $\phi_i\in [0,2\pi)$. The Killing vectors of the two $U(1)$ isometries are $K_\pm = \pm \tilde \ell^{-1} \p_{\varphi_1}+\ell^{-1} \p_{\varphi_2}$. Details on the structure of supersymmetric field theories on three-dimensional curved manifolds can be found in \cite{Closset:2012ru} (see also \cite{Benini:2013yva,Aprile:2016gvn}, where the special case of the ellipsoid is worked out in detail). The supersymmetric defects in this paper are wrapping the orbits of $K_+$. These orbits are periodic when $b^2=\frac{\tilde \ell}{\ell}$ is a rational number. Otherwise, they do not close and instead fill out the $(\varphi_1,\varphi_2)$-torus densely. In what follows, we focus on the case of $b^2$ being rational (the round $S^3$ has $b=1$ and is such a case), and $\omega$ is related to the dimensional radius $\mathfrak R$ of the closed orbit of $K_+$ by the relation $\mathfrak R = (2\pi \omega)^{-1} \sqrt{\ell \tilde \ell}$. \label{foot}} these can also be evaluated using supersymmetric localisation. 

Let us first consider building defect theories with 1D $\NN=\mathrm{2B}$ supersymmetry. The basic multiplets in such SQM theories are $\NN=\mathrm{2B}$ vector multiplets, chiral multiplets and Fermi multiplets (we refer the reader to \cite{Hori:2014tda} for a review). The one-loop contributions of each of these multiplets to the Witten index of 1D $\NN=\mathrm{2B}$ SQM theories are
\begin{align}\label{1DPFs}
  \begin{split}
g^{\mathcal{N}=2}_{\rm{vec}}(\lambda) =& \prod_{\beta\in \rm{Adj}} \widehat{\Gamma}_1\left(\langle\beta, \lambda \rangle\big| \omega\right)^{-1}\Gamma_1\left(\omega-\langle\beta, \lambda \rangle\big| \omega\right)^{-1} \cr 
=&\,\Gamma_1\left(\omega\big| \omega\right)^{-2\times \rm{rank}(\mathfrak g)}\prod_{\beta\in \rm{\Delta}} \widehat{\Gamma}_1\left(\langle\beta, \lambda \rangle\big| \omega\right)^{-1}\Gamma_1\left(\omega-\langle\beta, \lambda \rangle\big| \omega\right)^{-1}\;, \cr 
g^{\mathcal{N}=2}_{\rm{chi}}(\lambda, r) =& \prod_{\beta\in \mathcal{R}} \Gamma_1\left(\langle\beta, \lambda \rangle + \frac{r}{2}z\Big| \omega\right)\Gamma_1\left(\omega-\langle\beta, \lambda \rangle-  \frac{r}{2}z\Big| \omega\right)\;,\cr 
g^{\mathcal{N}=2}_{\rm{fer}}(\lambda, r) =& \prod_{\beta\in \mathcal{R}} \Gamma_1\left(-\langle\beta, \lambda \rangle - \frac{r}{2}z\Big| \omega\right)^{-1}\Gamma_1\left(\omega+\langle\beta, \lambda \rangle +  \frac{r}{2}z\Big| \omega\right)^{-1}\;,
\end{split}
\end{align} 
where $\mathcal{R}$ is a generic representation in any product of gauge or global symmetries and $\lambda$ is associated with a set of chemical potentials corresponding to these symmetries. As part of the global symmetries, we will always include an overall $U(1)_c$ that rotates the fundamental and anti-fundamental fields in the opposite way, as well as a $U(1)_d$ that acts on the adjoint fields. In addition, we have a generic $U(1)_R$ charge $r$ for chiral and Fermi multiplets, along with its chemical potential $z$. The $\Gamma_1(x|\omega)$ are Barnes 1-Gamma functions, defined in App.~\ref{math}.

Note that we can combine the Haar measure $\Delta^{\rm{Haar}}(\sigma)$ with the vector multiplet to obtain 
\begin{align}
  \Delta^{\rm{Haar}}(\lambda) g^{\mathcal{N}=2}_{\rm{vec}}(\lambda)=\Gamma_1\left(\omega\big| \omega\right)^{-2\times \rm{rank}(\mathfrak g)}\prod_{\beta\in \rm{\Delta}} \Gamma_1\left(\langle\beta, \lambda \rangle\big| \omega\right)^{-1}\Gamma_1\left(\omega-\langle\beta, \lambda \rangle\big| \omega\right)^{-1}\;.
\end{align}

With these ingredients we can also easily build  models with $\mathcal{N}=\mathrm{4A}$ supersymmetry. For this we simply need to use that a 1D $\mathcal{N}=4$ vector multiplet is a combination of an $\mathcal{N}=2$ vector multiplet and an adjoint $\mathcal{N}=2$ chiral multiplet with $r = 2$. Similarly, an $\mathcal{N}=4$ chiral multiplet of charge $r$ comprises of an $\mathcal{N}=2$ chiral multiplet with charge $r$ and an $\mathcal{N}=2$ Fermi multiplet with charge $(r-2)$, in the same representations of the gauge or global symmetry algebras.  

Combining this information with an identity given in the appendix, Eq.~\eqref{mathal}, we deduce that the contributions of $\NN=\mathrm{4A}$ vector and chiral multiplets to the SQM index can be succinctly written as 
\begin{align}\label{eq: 1DN=4 parts}
\begin{split}
\Delta^{\rm{Haar}}(\lambda) g^{\mathcal{N}=4}_{\rm{vec}}(\lambda) =& \left(\frac{\Gamma_1(z|\omega) \Gamma_1(\omega- z |\omega)}{\Gamma_1(\omega|\omega)^2}\right)^{\rm{rank}(\mathfrak g)}\prod_{\beta \in \Delta} \Delta_h \left(\langle \beta, \lambda\rangle \big| \omega, z\right)^{-1}\;, \cr
g^{\mathcal{N}=4}_{\rm{chi}}(\lambda, r) = &\prod_{\beta \in \mathcal{R}} \Delta_h \left(\langle\beta, \lambda\rangle + \frac{r}{2}z\Big| \omega, z\right)\;,
\end{split}
\end{align}
where the functions $\Delta_h$ and some of their properties are also given in App.~\ref{math}.

Post deconstruction the above will lift to the elliptic-genus contributions of 2D surface defects with $(2,2)$ and $(4,4)$ supersymmetry respectively. As we will see shortly, these results will be dominated by a closely-related function, $\Delta_e$, which will be related to $\Delta_h$ by an identity analogous to the one relating $\Gamma_h$ and $\Gamma_e$ in \eqref{mathagmain}.

\subsection{2D $\NN=(4,4)$ Defects in 4D $\NN=4$ SYM}\label{Neq4}

\begin{figure}
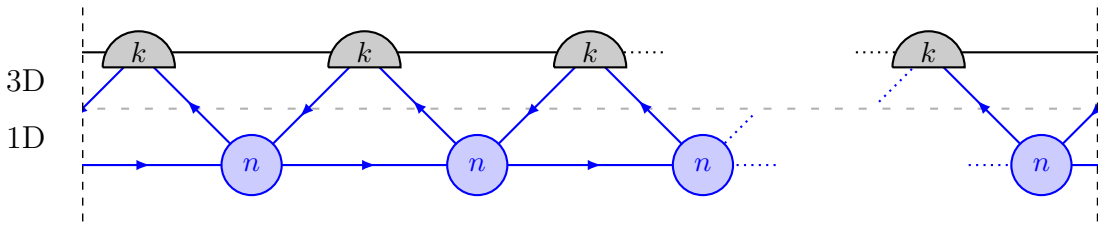

\begin{center}
	\threeDdefquivone
\end{center}
\caption{The 3D $\mathcal{N}=4$ circular quiver theory in the presence
  of 1D $\mathcal{N}=4$ defects. Black links and nodes are 3D
  $\mathcal{N}=4$ hypermultiplets and 3D $\mathcal{N}=4$ vector
  multiplets respectively of the bulk theory. The blue oriented links
  and nodes are 1D $\mathcal{N}=4$ chiral multiplets with assigned R charge $r=0$ and 1D $\mathcal{N}=4$ vector multiplets respectively for the defect theory. The diagonal 1D chiral multiplets have been assigned $U(1)_c$ charge $q_c=1$, while the horizontal ones $U(1)_d$ charge $q_d=1$.}\label{3DN8Orb}
\end{figure}

For this example, the starting point is an orbifold of a 3D $\mathcal{N}=8$ theory in the presence of a 1D $\mathcal{N}=8$ defect. In the language of the brane system of Tab.~\ref{IIB}, the defects are described by $nN$ D1 branes suspended between the $kN $ D3s and an NS5 in the $x^6$ direction. The $x^3$ direction is compactified and the branes are in the presence of the $A_{N-1}$ orbifold singularity. The quiver-gauge theory emerging at low energies is given by Fig.~\ref{3DN8Orb} and involves a 3D $\mathcal N=4$ bulk quiver with $U(k)$ gauge group nodes coupled to $\mathcal N=4$ 1D defects with $U(n)$ groups.\footnote{A class of closely-related systems with the same amount of supersymmetry appear in \cite{Ito:2016fpl}. Our brane configurations before deconstruction are related to those appearing in that reference by T duality and the low-energy quiver-gauge theories by dimensional reduction.}

The partition function of the combined 3D-1D system can be split into two parts as per
\begin{align}
\mathcal{Z}_{\rm{3D-1D}} =  \prod_{\alpha}\frac{1}{k!}\int \prod_{b=1}^k {\rm d}\sigma^{(\alpha)}_b \mathcal{Z}_{\rm 3D}(\sigma_b^{(\alpha)})\mathcal{Z}_{\rm 1D}(\sigma_b^{(\alpha)}) \;.
\end{align}
The bulk partition function contains the ingredients discussed in Sec.~\ref{pf} and for the quiver of Fig.~\ref{3DN8Orb} is given by
\begin{align}
\begin{split}
\mathcal{Z}_{\rm{3D}}(\sigma_b^{(\alpha)}) &= \prod_{\alpha}\widehat{\Gamma}_h \left(0\big| \omega_1, \omega_2\right)^{-k}\prod_{b\neq c}^k \Gamma_h \left(\sigma_b^{(\alpha)} - \sigma_c^{(\alpha)}\big| \omega_1, \omega_2\right)^{-1}\cr 
\times \prod_{b,c}^k& \Gamma_h \left(\frac{\omega_+}{2} - \sigma_b^{(\alpha +1)} + \sigma_c^{(\alpha)}\big| \omega_1, \omega_2\right) \Gamma_h \left(\frac{\omega_+}{2} +\sigma_b^{(\alpha +1)} - \sigma_c^{(\alpha)}\big| \omega_1, \omega_2\right)\;.
\end{split}
\end{align}

The defect contribution can be evaluated from the ingredients of Sec.~\ref{defingr} and is in turn given by 
\begin{align}\label{N4predec}
\begin{split}
&\mathcal{Z}_{{\rm 1D}}(\sigma_b^{(\alpha)})=  \prod_{\alpha}\frac{1}{n!} \left(\frac{\Gamma_1(z|\omega) \Gamma_1(\omega- z |\omega)}{\Gamma_1(\omega|\omega)^2}\right)^{k}  \oint  \prod_{j=1}^n{ \rm d} u_j^{(\alpha)}\frac{\prod_{i,j}^n \Delta_h \left(u^{(\alpha +1)}_i - u^{(\alpha)}_j +\kappa    |\omega , z \right)}{\prod_{i\neq j}^n \Delta_h \left(u^{(\alpha)}_i - u^{(\alpha)}_j|\omega , z \right)}\cr 
&\times \prod_{i=1}^n\prod_{b=1}^k \Delta_h \left(u_j^{(\alpha)} - \sigma_b^{(\alpha)}+  l|\omega, z \right)  \Delta_h \left( \sigma_b^{(\alpha)} -u_j^{(\alpha+1)}+  l |\omega, z \right) \;.
\end{split}
\end{align}
Note that this involves integrating the 1D gauge-group parameters $u_j^{(\alpha)}$ over some contour. These integrals can be eventually performed using the Jeffrey--Kirwan residue prescription---as e.g.\ in \cite{Benini:2013xpa,Assel:2015oxa}---although we will not do so in this paper.\footnote{The Jeffrey--Kirwan residue prescription is sensitive to the sign of a Fayet--Iliopoulos (FI) parameter in the SQM. We will implicitly assume that all the 1D gauge theories that appear in this paper are deformed by an FI term of definite sign. } Furthermore, for each defect node there is a $U(1)_c$ symmetry that rotates the bifundamental chiral multiplets connecting to the bulk theory, with associated chemical potential $l$, as well as a different $U(1)_d$ symmetry that acts on the bifundamental chiral multiplets between defect gauge nodes---associated with a chemical potential $\kappa$.  We have chosen these charges appropriately, $q_c = 1$ and $q_d = 1$, so as to deconstruct a superconformal 4D-2D system.

The exact-deconstruction procedure can now be applied to the 3D piece as in Sec.~\ref{pf}. It is implemented in a similar fashion for the 1D piece, namely, we identify the parameters $u_i^{(\alpha)}\to u_i$ and shift all the arguments of the functions appearing in \eqref{N4predec} by $m_\alpha = i\frac{\alpha}{R}$. Upon performing this operation $\mathcal{Z}_{\rm 3D-1D}$ becomes
\begin{align}
\mathcal{Z}^{\rm Dec}_{\rm{3D-1D}} =\frac{1}{k!} \oint \prod_b^k \frac{{\rm d} v_b}{2\pi i v_b} \mathcal{Z}^{\rm Dec}_{\mathrm{3D}}(v)\mathcal{Z}^{\rm Dec}_{\mathrm{1D}}(v) \;,
\end{align}
with 
\begin{align}
\mathcal{Z}^{\rm Dec}_{\mathrm{3D}}(v) = (\mathfrak{p};\mathfrak{p})^k (\mathfrak{q};\mathfrak{q})^k  \frac{\prod_{b,c}^k\Gamma_e(v_b v_c^{-1} (\mathfrak{p}\mathfrak{q})^{\frac{1}{4}} | \mathfrak{p},\mathfrak{q})^2}{\prod_{b\neq c}^k\Gamma_e\left(v_b v_c^{-1}\,|\, \mathfrak{p}, \mathfrak{q}\right)}\;,
\end{align}
which agrees with the index for $\mathcal{N}=4$ SYM as found in
\cite{Kinney:2005ej}.\footnote{This can be seen if in the notation of \cite{Kinney:2005ej} one considers the limit $w \to t^{\frac{1}{2}}$ and by mapping $v \to t^{-\frac{1}{2}}$, with $t = (\mathfrak{p}\mathfrak{q})^{\frac{1}{6}}$ and $y = \left(\mathfrak{p}\mathfrak{q}^{-1}\right)^{\frac{1}{2}}$, after the inclusion of the Haar measure.} In turn, the defect part becomes  
\begin{align}\label{1DN4}
\begin{split}
\mathcal{Z}^{\rm Dec}_{\rm 1D}(v)= & \left(\frac{(q;q)^2}{\theta(y^{-1}|q)}\right)^{k}  \oint  \prod_{j}^n\frac{{ \rm d} \mathfrak{u}_j}{2\pi i \mathfrak{u}_j} \frac{\prod_{i,j}^n \Delta_e \left(d\, \mathfrak{u}_i \mathfrak{u}^{-1}_j|q , y^{-1} \right)}{\prod_{i\neq j}^n \Delta_e  \left(\mathfrak{u}_i \mathfrak{u}^{-1}_j|q , y^{-1} \right)}\cr
&\times \prod_{j}^n\prod_{b}^k \Delta_e\left(\mathfrak{u}_j v_b^{-1}c|q, y^{-1} \right)  \Delta_e  \left(\mathfrak{u}_j^{-1} v_b \, c|q, y^{-1} \right) \;,
\end{split}
\end{align}
with $q = e^{-2 \pi R \omega}$, $y = e^{-2 \pi  R z}$, $c = e^{- 2 \pi  R l}$ and $d = e^{- 2
  \pi  R \kappa}$. Here we made use of the identity
\begin{align}
  \label{eq:10}
  \Delta_e(A | \Omega,T) = \prod_{\alpha=-\infty}^\infty \Delta_h \left( a + i \frac{\alpha}{R} \Big | \omega, t \right)~,
\end{align}
where $A = e^{-2\pi   Ra }$, $ \Omega = e^{-2\pi R\omega }$, $T= e^{-2\pi   Rt}$. The function $\Delta_e(A | \Omega, T)$ is defined as the ratio of theta functions
\beq
\label{mathanmain}
\Delta_e(A | \Omega, T) \equiv \frac{\theta(A T ; \Omega)}{\theta(A; \Omega)}
~.
\eeq

\begin{figure}[t]
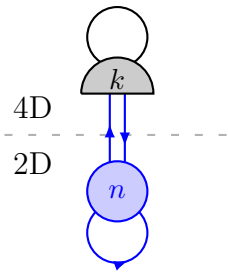

\begin{center}
	\fourDdefquivone
\end{center}
\caption{The 4D $\mathcal{N}=4$ SYM theory in the presence of a 2D $\mathcal{N}=(4,4)$ surface defect. Black unoriented links and nodes are 4D $\mathcal{N}=2$ hypermultiplets and 4D $\mathcal{N}=2$ vector multiplets respectively of the bulk theory. The blue oriented links and nodes are 2D $\mathcal{N}=(2,2)$ chiral multiplets and 2D $\mathcal{N}=(2,2)$ vector multiplets respectively for the defect theory.}\label{4DN4d}
\end{figure}

We would like to identify the above result as the index of a 2D $(4,4)$ theory coupled to the bulk 4D $\mathcal{N}=4$ SYM theory (see Fig.\ \ref{4DN4d}). For that purpose it is useful to recall some facts regarding elliptic genera in two-dimensional theories.\footnote{For a summary of the relevant details we refer the reader to \cite{Cordova:2017ohl}.} On the one hand, when considered as a partition function on $T^2$, e.g.\ such as in the supersymmetric-localisation calculations of \cite{Benini:2013nda,Benini:2013xpa}, the elliptic genus involves tracing over states in the R-R sector of the 2D (2,2) theory; these theories need not necessarily be superconformal. On the other hand, for superconformal theories,  the state-operator map exchanges R and NS boundary conditions and therefore when viewed as a generating function for operator counting it is more appropriately defined by tracing over the NS-NS sector, e.g. such as in the superconformal index calculations of \cite{Gadde:2013ftv,Cordova:2017ohl}.

The exact-deconstruction procedure in general takes an $\mathcal N=\mathrm{2B}$ SQM partition function on $S^1$ and lifts it to a (2,2) partition function on $T^2$. As such, one would expect it to reproduce the R-R elliptic genus. Indeed, in \eqref{1DN4} one recognises the elliptic-genus contributions attributed to a 2D (2,2) vector, an adjoint chiral (first line) and two (anti)fundamental chirals (second line) as given in \cite{Benini:2013xpa}. When combined with the bulk piece they make up the contributions attributed to the 4D-2D system summarised by the quiver of Fig.~\ref{4DN4d}.

The R-R and NS-NS elliptic genera can be related by spectral flow through the relation \begin{align}
  \label{flow}
  \mathcal G_{\mathrm{R-R}}(q,y) =  y^{c/6} \mathcal G_{\mathrm{NS-NS}}(q,q^{\frac{1}{2}}y) \;.
\end{align}
Implementing this dictionary in \eqref{1DN4} gives rise, up to an overall $c$-dependent coefficient, to 
\begin{align}\label{1DN4v2}
\begin{split}
\widehat{\mathcal{Z}}^{\rm Dec}_{\rm 1D}(v)= & \left(\frac{(q;q)^2}{\theta(q^{\frac{1}{2}}y^{-1}|q)}\right)^{k}  \oint  \prod_{j}^n\frac{{ \rm d} \mathfrak{u}_j}{2\pi i \mathfrak{u}_j} \frac{\prod_{i,j}^n \Delta_e \left(d\, \mathfrak{u}_i \mathfrak{u}^{-1}_j|q ,q^{\frac{1}{2}} y^{-1} \right)}{\prod_{i\neq j}^n \Delta_e  \left(\mathfrak{u}_i \mathfrak{u}^{-1}_j|q ,q^{\frac{1}{2}} y^{-1} \right)}\cr
&\times \prod_{j}^n\prod_{b}^k \Delta_e\left(\mathfrak{u}_j v_b^{-1}c|q, q^{\frac{1}{2}}y^{-1} \right)  \Delta_e  \left(\mathfrak{u}_j^{-1} v_b \, c|q,q^{\frac{1}{2}} y^{-1} \right) \;.
\end{split}
\end{align}
The identifications between defect and bulk fugacities 
\begin{align}
  \label{eq:12}
  q = \fq\;,\qquad y= \fp^{-\frac{1}{2}}\;,\qquad c = \fp^{-\frac{3}{8}} \fq^{\frac{1}{8}}\;,\qquad d = \fp^{-1}
\end{align}
lead to the final result 
\begin{align}
  \label{eq:13}
  \begin{split}
\widehat{\mathcal{Z}}^{\rm Dec}_{\rm 1D}(v)= & \left(\frac{(\fq;\fq)^2}{\theta(\sqrt{\fp \fq}|\fq)}\right)^{k}  \oint  \prod_{j}^n\frac{{ \rm d} \mathfrak{u}_j}{2\pi i \mathfrak{u}_j} \frac{\prod_{i,j}^n \Delta_e \left(\fp^{-1}\, \mathfrak{u}_i \mathfrak{u}^{-1}_j|\fq , \sqrt{ \fp \fq} \right)}{\prod_{i\neq j}^n \Delta_e  \left(\mathfrak{u}_i \mathfrak{u}^{-1}_j|\fq , \sqrt{ \fp \fq} \right)}\cr
&\times \prod_{j}^n\prod_{b}^k \Delta_e\left(\mathfrak{u}_j v_b^{-1}  \fp^{-\frac{3}{8}} \fq^{\frac{1}{8}}|\fq, \sqrt{ \fp \fq} \right)  \Delta_e  \left(\mathfrak{u}_j^{-1} v_b \,  \fp^{-\frac{3}{8}} \fq^{\frac{1}{8}}|\fq,\sqrt{ \fp \fq} \right) \;.  
\end{split}
\end{align}
This is precisely the answer for the $\mathcal N=1$ limit ($\ft\to\sqrt{\fp\fq}$) of the $\mathcal N=2$ superconformal index of the 4D-2D system of Fig.~\ref{4DN4d}, as calculated in \cite{Gadde:2013ftv}. Once again, one needs to perform the 2D gauge integrals using the Jeffrey--Kirwan residue prescription, which can be carried out explicitly as in \cite{Gadde:2013ftv,Benini:2013xpa}.

We can motivate the fugacity relations \eqref{eq:12} as follows. On the one hand, they are precisely the identifications used in the 4D-2D index calculation of \cite{Gadde:2013ftv}, where $q,y,c,d$ are fugacities for a 2D (2,2) index in the NS-NS sector.\footnote{For $q,y = q^{\frac{1}{2}} t^{-1}$ these relations can be found in Eq.~(5.31) of \cite{Gadde:2013ftv} by considering the limit $\ft\to \sqrt{\fp\fq}$, while for $c, d$ in their subsequent discussion.} On the other, they can also be derived from the specifics of the embedding of the defect SQM algebra into the 3D bulk superalgebra. The full details of the embedding in the case of the round $S^3$ can be found in \cite{Assel:2015oxa}. The general case of the ellipsoid can be worked out in a similar fashion.\footnote{As we pointed out in Fn.~\ref{foot}, in this section we are wrapping the defects along the orbits of the Hopf fibre. The flow of the Killing vector $K_+$ generates the isometry along this circle, which is therefore associated simultaneously with the fugacities $q$ and $\mathfrak q$, or equivalently the chemical potentials $i\omega$ and $\omega_2$. The remaining relations can be determined as in \cite{Assel:2015oxa}, by embedding an appropriate deformation of the 1D supersymmetry algebra $\{Q, \bar Q\} = H $---obtained by turning on Wilson lines/real mass parameters along the lines of\cite{Hori:2014tda}---into the 3D $\mathcal N=2 $ supersymmetry algebra on $S^3_b$, which reads $\{q,\bar q \} = -2 i K_+^\mu\partial_\mu +  \frac{2i}{\sqrt{\ell\tilde \ell}}r \omega_+$ \cite{Aprile:2016gvn}.}

\subsection{2D $\NN=(2,2)$ Defects in 4D $\NN=2$ SQCD}

Our next example is more general in that it involves half of the supersymmetry of the previous subsection and non-trivial flavour symmetries. The starting point is a brane system consisting of $kN$ D3s stretched between two NS5 branes along the interval $L_3$, with $N_1 N$ semi-infinite D3s extending to the left and $N_2 N$ to the right. Once again, there are also $n N$ D1s suspended between the rightmost NS5 and an additional NS5 in the $x^6$ direction and the whole system is probing the $A_{N-1}$ orbifold singularity. The corresponding 3D-1D quiver emerging at low energies is depicted in Fig.~\ref{exampletwo}.\footnote{There also exists a dual UV description of the same vortex loop, obtained by having the D1s end on the leftmost NS5 along $x^3$. This leads to a quiver with $N_1$ and $N_2$ exchanged and the opposite sign for the 1D FI term compared to the system we are using. This sign difference is crucial for recovering the same partition function from both configurations. For a detailed account see \cite{Assel:2015oxa}.}

From the 1D point of view, the fundamental/antifundamental chiral and Fermi multiplets have unit charge under the $U(1)_c$ symmetry. The adjoint chiral multiplets and bifundamental Fermi multiplets have unit charge under the $U(1)_d$ symmetry.

\begin{figure}[t]
\begin{center}
\hspace*{-1.3cm}	\drawquiverdefextwo
\end{center}
\caption{The 3D $\mathcal{N}=2$ circular-quiver theory in the presence
  of 1D $\mathcal{N}=2$ defects. Black oriented links and nodes are 3D
  $\mathcal{N}=2$ chiral multiplets and 3D $\mathcal{N}=2$ vector
  multiplets respectively of the bulk theory. The blue oriented links
  and nodes are 1D $\mathcal{N}=2$ chiral multiplets with R charge $r=0$ and 1D $\mathcal{N}=2$ vector multiplets respectively for the defect theory. The orange oriented links are $\NN = 2$ Fermi multiplets for the defect theory. The diagonal 1D chiral multiplets have been assigned $U(1)_c$ charge $q_c=1$, while the horizontal ones $U(1)_d$ charge $q_d=1$.}\label{exampletwo}
\end{figure}

As before, the $S^3_b$ partition function can be split into 3D and 1D contributions: 
\begin{align}
\mathcal{Z}_{\rm{3D-1D}} =  \prod_{\alpha}\frac{1}{k!}\int \prod_{b=1}^k {\rm d}\sigma^{(\alpha)}_b \mathcal{Z}_{\mathrm{3D}}(\sigma_b^{(\alpha)})\mathcal{Z}_{\mathrm{1D}}(\sigma_b^{(\alpha)}) \;.
\end{align}
The 3D part is 
\begin{align}
\begin{split}
&\mathcal Z_{\mathrm{3D}}(\sigma^{(\alpha)}) =\prod_{\alpha} \Delta^{\text{Haar}} \left(\sigma^{(\alpha)}\right) \prod_{b,c=1}^{k} \frac{\Gamma_h\left(\omega_+ + \sigma^{(\alpha)}_b - \sigma^{(\alpha+1)}_c \Big| \omega_1, \omega_2\right)}{ \widehat{\Gamma}_h\left(\sigma^{(\alpha)}_b - \sigma^{(\alpha)}_c \Big| \omega_1, \omega_2\right) }\\
                                                        &\times \prod_{b=1 }^k\prod_{j=1}^{N_1} \Gamma_h\left(\frac{1}{2}\omega_+ - \mu^{(\alpha)}_j +\sigma^{(\alpha)}_b    \Big| \omega_1, \omega_2\right) \Gamma_h\left(\frac{1}{2}\omega_+ - \sigma^{(\alpha +1 )}_b +  \mu_j^{(\alpha)} \Big| \omega_1, \omega_2\right) \\ 
                                                        &\times \prod_{b=1 }^k\prod_{m=1}^{N_2} \Gamma_h\left(\frac{1}{2}\omega_+ -\sigma^{(\alpha +1)}_b+    \nu^{(\alpha+1)}_m  \Big| \omega_1, \omega_2\right) \Gamma_h\left(\frac{1}{2}\omega_+ - \nu^{(\alpha +1 )}_m +  \sigma_b^{(\alpha)} \Big| \omega_1, \omega_2\right)\;,
\end{split}
\end{align}
where $\mu_j^{(\alpha)}, \nu_m^{(\alpha)}$ are new chemical potentials associated with the global symmetries, while the other chemical potentials were defined in Sec.~\ref{Neq4}. The 1D part is in turn
\begin{align}
\begin{split}
&\mathcal{Z}_{{\rm 1D}}(\sigma^{(\alpha)})=  \cr &  \oint \prod_{\alpha}\frac{1}{n!} \Delta^{\text{Haar}} \left(u^{(\alpha)}\right)  \prod_{j}{ \rm d} u_j^{(\alpha)}\prod_{i,j=1}^n\frac{ \Gamma_1 \left(u_i^{(\alpha)} - u^{(\alpha+1)}_j \big| \omega\right)\Gamma_1 \left(\omega - u_i^{(\alpha)} + u^{(\alpha+1)}_j \big| \omega\right)}{\widehat{\Gamma}_1 \left(u_i^{(\alpha)} - u_j^{(\alpha)}\big|\omega\right) \Gamma_1 \left(\omega + u_i^{(\alpha)} - u_j^{(\alpha)}\big|\omega\right)}\cr 
\times & \prod_{i,j=1}^n\frac{ \Gamma_1 \left(u_i^{(\alpha)} - u^{(\alpha)}_j  + \kappa\big| \omega\right)\Gamma_1 \left(\omega - u_i^{(\alpha)} + u^{(\alpha)}_j -\kappa \big| \omega\right)}{ \Gamma_1 \left(-u_i^{(\alpha)} + u^{(\alpha+1)}_j +z -\kappa\big| \omega\right)\Gamma_1\left(\omega + u_i^{(\alpha)} - u^{(\alpha+1)}_j -z + \kappa\big| \omega\right)}\cr 
\times & \prod_{i=1}^n \prod_{b=1}^k   \frac{ \Gamma_1 \left(u_i^{(\alpha)} - \sigma^{(\alpha+1)}_b  +l\big| \omega\right)\Gamma_1 \left(\omega - u_i^{(\alpha)} + \sigma^{(\alpha+1)}_b  -l\big| \omega\right)}{ \Gamma_1 \left(-u_i^{(\alpha)} + \sigma^{(\alpha)}_b +z -l\big| \omega\right)\Gamma_1\left(\omega + u_i^{(\alpha)} - \sigma^{(\alpha)}_b-z +l\big| \omega\right)}\cr
\times &\prod_{i=1}^n \prod_{m=1}^{N_2}   \frac{ \Gamma_1 \left(-u_i^{(\alpha)} + \nu^{(\alpha)}_m  +l\big| \omega\right)\Gamma_1 \left(\omega + u_i^{(\alpha)} - \nu^{(\alpha)}_m -l\big| \omega\right)}{ \Gamma_1 \left(u_i^{(\alpha)} - \nu^{(\alpha+1)}_m +z -l\big| \omega\right)\Gamma_1\left(\omega - u_i^{(\alpha)} + \nu^{(\alpha+1)}_m -z+l\big| \omega\right)}\;.
\end{split}
\end{align}
 
Once again, in order to implement dimensional deconstruction at the level of partition functions, we break the relevant groups involved to a diagonal subgroup and shift the arguments of the Gamma functions, which results in a product over the KK mass parameters, $m_\alpha =i \frac{\alpha}{R}$. 

\begin{figure}
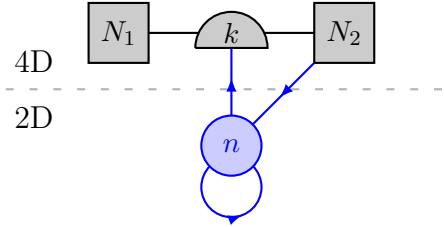
 
\begin{center} 
  \drawquiverdefexthreeres
\end{center}
\caption{The 4D $\mathcal{N}=2$ SYM theory in the presence of a 2D $\mathcal{N}=(2,2)$ surface defect. Black unoriented links and semi-circular nodes are 4D $\mathcal{N}=2$ hypermultiplets and 4D $\mathcal{N}=2$ vector multiplets respectively of the bulk theory. The black squares denote 4D flavour groups. The blue oriented links and nodes are 2D $\mathcal{N}=(2,2)$ chiral multiplets and 2D $\mathcal{N}=(2,2)$ vector multiplets, respectively, for the defect theory.}\label{defexampletwo} 
\end{figure}

The resultant expression is much simpler and when expressed in terms of fugacities reads
\begin{align}
\mathcal{Z}^{\rm Dec}_{\rm{3D-1D}} =\frac{1}{k!} \oint \prod_b \frac{{\rm d} v_b}{2\pi i v_b} \mathcal Z^{\textrm{Dec}}_{\rm{3D}}(v)\mathcal{Z}^{\rm Dec}_{\rm 1D}(v) \;,
\end{align}
with 
\begin{align}
\begin{split}
  \mathcal Z^{\textrm{Dec}}_{\mathrm{3D}}(v) = (\fp;\fp)^{k} (\fq;\fq)^{k} \prod_{b\neq c}\Gamma_e\left(v_bv_c^{-1}\Big|\fp,\fq\right)^{-1}\prod_{b=1 }^k\prod_{j=1}^{N_1} \Gamma_e\left((\fp\fq)^{\frac{1}{4}} (v_b s_j^{-1})^{\pm} \Big| \fp,\fq\right)\cr 
  \times\prod_{b=1 }^k\prod_{m=1}^{N_2} \Gamma_e\left((\fp\fq)^{\frac{1}{4}} (v_b t_m^{-1})^{\pm} \Big| \fp,\fq\right) \;,
  \end{split}
\end{align}
where $s_j = e^{-2 \pi i R \mu_j}$ and $t_m = e^{-2 \pi i R \nu_m}$. The defect piece becomes just
\begin{align}
  \begin{split}
    \mathcal{Z}^{\rm Dec}_{\rm 1D}(v)= & \left(\frac{(q;q)^2}{\theta(y^{-1}|q)}\right)^{k}  \oint  \prod_{j}^n\frac{{ \rm d} \mathfrak{u}_j}{2\pi i \mathfrak{u}_j} \frac{\prod_{i,j} \Delta_e  \left(d\, \mathfrak{u}_i \mathfrak{u}^{-1}_j |q ,y^{-1} \right)}{\prod_{i\neq j} \Delta_e  \left(\mathfrak{u}_i \mathfrak{u}^{-1}_j|q , y^{-1} \right)}\cr
&\times \prod_{j=1}^n\prod_{b=1}^k \Delta_e\left(\mathfrak{u}_j v_b^{-1}c\,|q, y^{-1} \right) \prod_{m=1}^{N_2} \Delta_e  \left(\mathfrak{u}_j^{-1} t_m \, c\,|q, y^{-1} \right) \;,
\end{split}
\end{align}
which upon making use of the identifications that realise the embedding of the defect into the bulk \eqref{eq:12} leads to
\begin{align}
\mathcal{Z}^{\rm Dec}_{\rm 1D}(z)= & \left(\frac{(\fq;\fq)^2}{\theta(\fp^{\frac{1}{2}}|\fq)}\right)^{k}  \oint  \prod^n_{j}\frac{{ \rm d} \mathfrak{u}_j}{2\pi i \mathfrak{u}_j} \frac{\prod_{i,j} \Delta_e  \left(\fp^{-1}\, \mathfrak{u}_i \mathfrak{u}^{-1}_j |\fq , \fp^{\frac{1}{2}} \right)}{\prod_{i\neq j} \Delta_e  \left(\mathfrak{u}_i \mathfrak{u}^{-1}_j|\fq , \fp^{\frac{1}{2}} \right)}\cr
  &\times \prod_{j=1}^n\prod_{b=1}^k \Delta_e\left(\mathfrak{u}_j v_b^{-1}\fp^{-\frac{3}{8}}\fq^{\frac{1}{8}}\,|\fq, \fp^{\frac{1}{2}} \right) \prod_{m=1}^{N_2} \Delta_e  \left(\mathfrak{u}_j^{-1} t_m \, \fp^{-\frac{3}{8}}\fq^{\frac{1}{8}}\,|\fq, \fp^{\frac{1}{2}} \right) \;.
\end{align}
This is the combined $S^3_b \times S^1_{\widehat R}$ partition function for the quiver depicted in Fig.~\ref{defexampletwo} in the R-R sector for the 1D defect partition function. In the case of $N_1 = N_2 = k$, one can once again implement the spectral flow argument \eqref{flow} to recover the corresponding index result in \cite{Gadde:2013ftv}.

We close this discussion by noting that, although we have examined two very specific examples involving vortex loops, one can clearly apply the method of exact deconstruction to more general setups. These can be engineered using branes through various combinations of the ingredients from Tab.~\ref{IIB}, including combinations of NS5 and NS5$'$ branes at different points along the $x^6$ direction connected with different numbers of D1s. As a result, one can obtain e.g.\ all examples of \cite{Assel:2015oxa} placed on an orbifold singularity, which will result in product defect gauge groups, additional matter at each quiver node and so on. The implementation of exact deconstruction to these theories is straightforward.

\section{4D Surface Defects from 3D Wilson Loops}
\label{wilson}

In Sec.~\ref{vortex} we demonstrated how to deconstruct codimension-two surface defects in 4D $\NN =4$/$\NN=2$ theories that preserve non-chiral 2D $\NN = (4,4)$/$\NN=(2,2)$ supersymmetry from vortex loops in 3D $\NN = 4$/$\NN=2$ theories. The vortex loops in three dimensions were defined by coupling 1D $\NN = 4$/$\NN=2$ SQM to the 3D bulk. Similarly, the deconstructed surface defects were defined by coupling 2D $\NN=(4,4)$/$\NN=(2,2)$ theories to the 4D bulk. 

Another prominent, and perhaps more common, class of line defects in 3D theories arises from Wilson loops. In 3D $\NN=4$ theories $\frac{1}{2}$-BPS Wilson-loop operators, which will be the focus of our discussion in the rest of this section, are given by path-ordered traces of the form
\beq
\label{wilsonaa}
W_\RR = \Tr_\RR P \exp \oint i \left( A_\mu \dot x^\mu + \sqrt{-\dot x^2} \sigma \right) d\tau
~.
\eeq
In this formula $\sigma$ is the real scalar field in the $\NN=2$ vector multiplet, which is part of the 3D $\NN=4$ vector multiplet. As usual, Wilson loops are labelled by a gauge-group representation $\RR$. For concreteness, in what follows we will focus on the fundamental and antisymmetric representations of the unitary gauge group (namely, representations labelled by Young tableaux with a single column). Analogous statements will apply to more general representations. As a one-dimensional defect, the operator \eqref{wilsonaa} preserves the 1D $\NN=4\mathrm{B}$ supersymmetry that arises from the dimensional reduction of the {\it chiral} 2D $\NN=(4,0)$ supersymmetry. This should be compared against the 2D $\NN=(2,2)$ supersymmetry associated by dimensional reduction with the vortex loops.

In supersymmetric localisation the insertion of a Wilson-loop operator \eqref{wilsonaa} in the round $S^3$ partition function\footnote{For the moment we consider the case of a round $S^3$. We will soon generalise to squashed three-spheres with arbitrary squashing.} is captured by the introduction of a factor
\beq
\label{wilsonab}
\Tr_\RR \left( e^{2\pi \sigma} \right) = \sum_{\beta \in \RR} e^{2\pi \langle \beta,\sigma\rangle }
\eeq
in the resultant matrix integral \cite{Pestun:2007rz,Kapustin:2009kz}. On the RHS of this expression the sum is performed over all the weights $\beta$ of the representation $\RR$.

The question of whether one can implement exact deconstruction with 3D Wilson-loop insertions is an obvious one. However, if there exists a well-defined procedure for lifting 3D Wilson loops in some representation $\RR$, what kind of defect does one expect to obtain in four dimensions? The natural answer seems to be a codimension-two defect with chiral supersymmetry labelled by the same representation.

The literature on surface defects with 2D $\NN=(4,0)$ supersymmetry is relatively-limited.\footnote{We note that the generating function for a class of BPS co-dimension two defects on $S^3\times S^1$ with $(2,0)$ supersymmetry was recently obtained from the affinisation of the $S^3$ partition function including Wilson-loop insertions, followed by a projection with affine characters \cite{Lodin:2017lrc,Nieri:2015dts}. This procedure produces results that are very similar to the ones given in this section through deconstruction. It would be very interesting to further explore the relationship between the two prescriptions.} An early discussion of surface defects with 2D $\NN=(8,0)$ supersymmetry appears in \cite{Buchbinder:2007ar} and is based on the D3-D7 intersection in string theory; see \cite{Harvey:2007ab} for related work. The surface defects in \cite{Buchbinder:2007ar} are formulated by integrating out 2D chiral fermions. Notice that a similar approach to Wilson loops in 4D $\NN=4$ SYM was employed in \cite{Gomis:2006sb,Gomis:2006im}. In that case the result of integrating out the chiral fermions is not a single Wilson loop, but rather a sum over Wilson loops in different representations; this is a point which we will come back to more explicitly in a moment. Here we stress that the codimension-two operators obtained in this manner are formulated in terms of a Wess--Zumino--Witten action supported on a surface, with no explicit reference to individual representations \cite{Buchbinder:2007ar}.

Applying deconstruction to 3D Wilson loops is therefore an interesting direction that has the potential to produce novel results about chiral surface operators in 4D theories. In this section, we take the first steps towards this direction by studying deconstruction at the level of $S^3_b$ partition functions with Wilson loop insertions.\footnote{Chiral surface defects in 4D $\NN=2$ or $\NN=4$ SYM theories can also be obtained from the 6D $\NN=(2,0)$ theory by dimensionally reducing codimension-two or codimension-four defects. Related brane constructions of such defects appear, for example, in \cite{Agarwal:2018tso,Tachikawa:2011dz}. These constructions produce surface defects that are naturally labelled by representations of the gauge group.}

The strategy is straightforward and begins with a 3D circular quiver-gauge theory with a Wilson-loop insertion for each node.  That in turn introduces a product of insertions of the type \eqref{wilsonab} into the $S^3_b$ partition function. Then, one has to implement the deconstruction procedure and evaluate the result in the appropriate limit. We have previously argued that in the context of $S_b^3$ partition functions the main effect of the exact-deconstruction procedure is to produce a shift in the argument of various Gamma functions. This encoded the reorganisation of the 3D spectrum into the KK modes of a 4D theory on a circle. However, from that point of view it is not immediately clear how we should extend the prescription to treat insertions of the type \eqref{wilsonab}. To handle this issue we instead propose the following approach.

First, it is convenient to take a step back and reformulate 3D Wilson loops in terms of a gauged 1D $\NN=2$ Fermi multiplet in analogy with \cite{Gomis:2006sb}---see \cite{Assel:2015oxa} for a related discussion in three dimensions.\footnote{The 3D $\mathcal N=2$ Fermi multiplet is also a representation of the 3D $\mathcal N=4$ supersymmetry algebra, and in that context it is sometimes called a half-Fermi multiplet; see e.g.\ the appendix of \cite{Lawrie:2016axq} for a 2D version.} For starters, consider a gauge theory with a single $U(k)$ node. We insert a one-dimensional defect described by a gauged $\NN=2$ Fermi multiplet in the fundamental representation of the bulk $U(k)$ gauge group. This can be engineered using the ingredients of Tab.~\ref{IIBW} in terms of $k$ D3s suspended between two NS5s along $x^3$ and a single D5$'$ separated by some distance in the $x^7$ direction. The fermion $\chi$ in the Fermi multiplet emerges from the quantisation of open strings connecting the D3s and D5$'$ and has the action
\beq
\label{wilsonac}
\int dt \, \chi^\dagger \Big[i \p_t + (A_0 + \frac{1}{\sqrt{\ell \tilde \ell}}\sigma- \frac{1}{\sqrt{\ell \tilde \ell}} i m)\Big] \chi
~,
\eeq
where $A_0$ is the temporal component of the bulk gauge field, $\sigma$ is the bulk Coulomb-branch parameter and $m$ is a mass parameter that can be viewed as the vev of a background $U(1)$ gauge field, corresponding to the frozen dynamics of the single D5$'$. Integrating out this fermion leads to the insertion of a $\frac{1}{2}$-BPS Wilson loop \cite{Gomis:2006sb,Assel:2015oxa}. Let us verify this statement at the level of the $S^3_b$ partition function. As in the vortex-loop case, we will take the Wilson-loop defect to wrap the Hopf fibre of the ellipsoid.

Up to an overall regularisation-dependent factor, the contribution of such a Fermi multiplet \eqref{1DPFs} can be re-expressed as 
\begin{align}
\begin{split}
\label{wilsonad}
g^{\mathcal N=2}_{\mathrm{fer}}(\sigma,m) &= \prod_{b=1}^k \sin \pi ( i \omega^{-1}\sigma_b + \omega^{-1} m ) \propto \prod_{b=1}^k \left( e^{-\pi\omega^{-1} \sigma_b + i \pi \omega^{-1} m} - e^{\pi \omega^{-1} \sigma_b - i \pi\omega^{-1} m} \right)
\\
&=e^{i\pi \omega^{-1} k m} e^{- \pi \omega^{-1} \sum_b \sigma_b}  \prod_{b=1}^k \left( 1- e^{-2\pi i \omega^{-1}  m} e^{2\pi \omega^{-1} \sigma_b} \right)
~.
\end{split}
\end{align}
The factor $e^{-\pi \omega^{-1}\sum_b \sigma_b}$ is related to a global anomaly for $U(1)\subset U(k)$ and is also noted in \cite{Assel:2015oxa}, where it is argued that it is cancelled by a bare supersymmetric Chern-Simons term in the bulk at level $\kappa=\frac{1}{2}$. Since
\beq
\label{wilsonae}
\prod_{b=1}^k \left( 1- e^{-2\pi i \omega^{-1} m} e^{2\pi\omega^{-1} \sigma_b} \right) = \sum_{l=0}^k (-1)^\rho{} e^{-2\pi i\omega^{-1} m \rho{}} \sum_{\beta\in \AA_\rho{}} e^{2\pi \omega^{-1} \langle\beta \cdot \sigma\rangle}
\eeq
we recover from \eqref{wilsonad} the expansion of the  Fermi-multiplet contribution  as a sum over all Wilson loops in the anti-symmetric representations $\AA_\rho$. From this result, the contribution of the $\rho$-antisymmetric representation can be recovered by noticing that it is weighted by the $\rho$-th power of the factor $\mu=e^{-2\pi i \omega^{-1} m}$ and hence isolated by evaluating the residue 
\beq
\label{wilsonaea}
\frac{1}{2\pi i} \oint d\mu \mu^{-\rho{}-1} g^{\mathcal N=2}_{\mathrm{fer}}(\sigma,\mu) ~.  \eeq

This observation motivates the following three-step approach for extending exact deconstruction to 3D Wilson loops:
\begin{enumerate}
\item[(i)] Add a 1D defect described by a Fermi multiplet for each node of the 3D quiver. 

  \item[(ii)] Use this to deconstruct a 4D theory with a chiral 2D surface defect described by a 2D Fermi multiplet.

    \item[(iii)] Isolate the contributions associated with different powers of $\mu$.
\end{enumerate}
The last point that needs to be addressed is whether the final step in the above prescription truly defines quantities that correspond to $k$ independent objects in 2D. The latter could in turn be interpreted as chiral surface operators labelled by antisymmetric representations of the four-dimensional bulk gauge group $U(k)$. To address this matter let us examine in detail the results obtained by this prescription.

For the first step, and as in any of the constructions of the previous sections our starting point is a 3D $N$-noded quiver, with each node labelled by $\alpha = \lfloor -\frac{N}{2}\rfloor +1, \ldots,\lfloor \frac{N}{2}\rfloor$. The specific details of the quiver are not important ---we will only assume that it is a 3D $\NN=2$ or $\NN = 4$ quiver that deconstructs to a 4D $\NN=2$ or $\NN=4$ gauge theory on a circle. For concreteness, one can consider engineering such an example by taking the brane system leading to \eqref{wilsonac} and placing it on the orbifold singularity of Tab.~\ref{IIBW}. At low energies, each node of the resultant quiver includes a 1D defect with chiral supersymmetry, described by an $\NN = 2$ Fermi multiplet of mass $m^{(\alpha)}$. This system is placed on $S^3_b$ and as already mentioned the 1D defect at each node wraps the Hopf fibre.

Before exact deconstruction, the total 1D contribution to the $S_b^3$ partition function is
\beq
\label{wilsonaf}
\mathcal Z_{\mathrm{1D}}(\sigma_b^{(\alpha)}, m_b^{(\alpha)})  = \prod_{\alpha} g^{\mathcal N=2}_{\mathrm{fer}}(\sigma^{(\alpha)}, m^{(\alpha)}) 
\eeq
with each Fermi-multiplet contribution given by\footnote{We have switched off all other chemical potentials that could appear here for the purposes of this discussion.}
\beq
\label{wilsonag}
g^{\mathcal N=2}_{\mathrm{fer}}(\sigma^{(\alpha)}, m^{(\alpha)})  = \prod_{b=1}^k \Gamma_1 \left( - \sigma^{(\alpha)}_{b}- m^{(\alpha)} \Big | \omega \right)^{-1} \Gamma_1\left( \omega +  \sigma_{b}^{(\alpha)} + m^{(\alpha)}  \Big | \omega \right)^{-1}
~.
\eeq
To deconstruct, we set $\sigma^{(\alpha)} \to \sigma$ and $m^{(\alpha)}\to m$, shift the arguments of the $\Gamma_1$ functions and take $N\to \infty$ to obtain 
\begin{align}
\begin{split}
\label{wilsonai}
\mathcal Z^{\mathrm{Dec}}_{\mathrm{1D}}(\sigma, m) &= \prod_{\alpha=-\infty}^\infty \prod_{b=1}^k \Gamma_1 \left( - \sigma_{b} - m + i\frac{\alpha}{R} \Big | \omega \right)^{-1} \Gamma_1\left( \omega +  \sigma_{b} + m +i\frac{\alpha}{R} \Big | \omega \right)^{-1}
\\
&\propto \prod_{\alpha=-\infty}^\infty \prod_{b=1}^k \prod_{n=-\infty}^\infty \left( -\sigma_b - m  +i\frac{\alpha}{R} + n \omega \right)
~.
\end{split}
\end{align}
Up to an unimportant zeta-function regularisable factor
\begin{align}
\begin{split}
\label{wilsonaj}
\mathcal Z^{\mathrm{Dec}}_{\mathrm{1D}}(\sigma, m) \propto \prod_{\alpha=-\infty}^\infty \prod_{b=1}^k \prod_{n=-\infty}^\infty \left( \omega^{-1} (-\sigma_b - m) + n + \alpha \tau \right)= \prod_{b=1}^k \theta \left( e^{- 2\pi i \omega^{-1}\sigma_b} \mu | q \right)
~,
\end{split}
\end{align}
where we defined $\tau = - \frac{i}{\omega R}$, $\mu = e^{-2\pi i \omega^{-1} m}$, $q = e^{2\pi i \tau}$ and $\theta$ is the theta function given in Eq.~\eqref{mathao}. This completes the second step of our prescription. 

For the third and final step we compute the quantity
\beq
\label{wilsonak}
\mathcal Z^{\mathrm{Dec} \; (\rho{})}_{\mathrm{1D}}(\sigma) \equiv \frac{1}{2\pi i} \oint d\mu \, \mu^{-\rho{}-1}
\prod_{b=1}^k \theta \left( e^{- 2\pi i \omega^{-1}\sigma_b} \mu | q \right)
~.
\eeq
Since $\theta(z|q)$ is closely related to the Jacobi-theta function,
\begin{align}
\begin{split}
\label{wilsonal}
\vartheta (z |q) =\prod_{n=1}^{\infty} (1-q^n) (1+ z q^{n-\frac{1}{2}}) \left(1+ z^{-1} q^{n-\frac{1}{2}} \right) = \sum_{n=-\infty}^\infty z^n q^{\frac{n^2}{2}}\;,
\end{split}
\end{align}
and the eta function, $\eta(q) = q^{\frac{1}{24}} \prod_{n=1}^\infty (1-q^n)$, via the relation
\beq
\label{wilsonam}
\theta(z|q) = q^{\frac{1}{24}} \eta^{-1}(q) \vartheta(-z q^{-\frac{1}{2}}|q )\;,
\eeq
we can re-write \eqref{wilsonak} as
\beq
\label{wilsonao}
\mathcal Z^{\mathrm{Dec} \; (\rho)}_{\mathrm{1D}}(\sigma) = (-1)^\rho q^{\frac{\rho}{2}+\frac{k}{24}} \, \eta^{-k}(q) \sum_{n_1,\ldots, n_k=-\infty}^\infty \delta_{\rho, \sum_b n_b} \, e^{-2\pi i  \omega^{-1} \sum_b n_b \sigma_b } q^{\frac{1}{2} \sum_{j=b}^k n_b^2}
~.
\eeq
Therefore, our proposal for the index of a chiral surface defect, labelled by an integer $\rho$, in 4D $\NN=2$ or $\NN=4$ theories is given by
\begin{align}
  \label{eq:14}
\mathcal I^{(\rho)}_{\mathrm{4D-2D}} = \oint \prod_{b=1}^k \frac{{\rm d} v_b}{2 \pi i v_b} \mathcal Z_{\rm{3D}}^{\rm{Dec}}(v)  \mathcal Z^{\mathrm{Dec} \; (\rho)}_{\mathrm{1D}}(v^\tau)\;,
\end{align}
with $v_b = e^{- 2 \pi i R \sigma_b}$. Now one can ask whether $\rho$ could be related to the $\rho$-antisymmetric representation in a suitable classification.

At first sight, the association with the fundamental or antisymmetric representations of $U(k)$ fails because the integer $\rho$ that appears in \eqref{wilsonao} is not restricted to $k$ different values. Nevertheless, we notice that $\mathcal Z^{\mathrm{Dec} \; (\rho)}_{\mathrm{1D}}(\sigma)$ as defined in \eqref{wilsonao} obeys the following quasi-periodicity relation 
\beq
\label{wilsonap}
\mathcal Z^{\mathrm{Dec} \; (\rho{}+k)}_{\mathrm{1D}}(\sigma) = (-1)^k q^{k+\rho} e^{-2\pi i\omega^{-1} \sum_b \sigma_b} \, \mathcal Z^{\mathrm{Dec} \; (\rho)}_{\mathrm{1D}}(\sigma)
~,
\eeq
which is easily derived by shifting all sums over $n_b$ by one unit.

The $\sigma$-dependent factor $e^{-2\pi i\omega^{-1} \sum_b \sigma_b}$ would cancel if the definition \eqref{wilsonao} included an extra $e^{2\pi i\omega^{-1}\frac{\rho}{k} \, \sum_b \sigma_b}$,\footnote{We remind the reader that a similar $e^{-\pi \sum_b \sigma_b}$ factor had to be cancelled in \eqref{wilsonad} for the case of the round $S^3$.} or if the bulk gauge group was $SU(k)$. With this cautionary note in mind,  the deconstructed 3D-1D squashed $S^3$ partition function is restricted to $k$ different values that could eventually fit into a classification of chiral surface defects in terms of the fundamental or antisymmetric representations of the gauge group. 

As a brief illustration we consider the case of a $U(2)$ gauge group, i.e.\ $k=2$. Then, it is straightforward to show that $\ZZ_{\mathrm{1D}}^{\mathrm{Dec} \, (\rho)}$ in \eqref{wilsonao} only assumes two different values\footnote{This is up to an overall $e^{-\pi \omega^{-1} \rho \sum_b \sigma_b}$ factor, which can be cancelled by the $e^{2\pi i \omega^{-1}\frac{\rho}{k} \sum_b \sigma_b}$ factor proposed in the previous paragraph.}
\begin{align}
\begin{split}
\label{wilsonaq}
e^{\pi i\omega^{-1} \rho\, \sum_b \sigma_b} q^{-\frac{\rho^2}{4}} \,  \mathcal Z^{\mathrm{Dec} \; (\rho = {\rm even})}_{\mathrm{1D}}(\sigma) &= 
\vartheta_{00}\left( e^{-2\pi i \omega^{-1} (\sigma_1-\sigma_2)} \Big | q^2 \right)
~,
\\
e^{\pi i\omega^{-1} \rho\, \sum_b \sigma_b} q^{-\frac{\rho^2}{4}} \,  \mathcal Z^{\mathrm{Dec} \; (\rho = {\rm odd})}_{\mathrm{1D}}(\sigma) &= 
\vartheta_{10}\left( e^{-2\pi i \omega^{-1} (\sigma_1-\sigma_2)} \Big | q^2 \right)
~,
\end{split}
\end{align}
where $\vartheta_{00}, \vartheta_{10}$ are the standard theta-function variants with characteristics. It is possible to manipulate the restricted sum in \eqref{wilsonao} for general $k$ and express it in terms of the product of $k-1$ theta functions, but unfortunately we have not uncovered simple expressions like the ones in \eqref{wilsonaq} for general $k$.

As in the vortex-loop case, we should mention that more complicated Wilson-loop insertions can be used as the starting point for deconstruction, where the associated representations of the bulk gauge group can be symmetric, antisymmetric or products thereof. The low-energy quivers can be obtained by generalising the brane system introduced above \eqref{wilsonac} to include combinations of D5 and D5$'$ branes at different points along the $x^7$ direction and connected with fundamental strings, always in the presence of the orbifold singularity. A detailed account of these constructions in the absence of an orbifold can be found in \cite{Assel:2015oxa}.

\section{Deconstruction of Codimension-one Defects}\label{lines}

We now switch gears from codimension-two to codimension-one defects. Deconstruction can be straightforwardly extended to circular quivers with localised modifications, which naturally lead to codimension-one defects in the emerging higher-dimensional theory. From the point of view of the SQM on the deconstructed $S^1$, these defects are local operators inserted at different points of the circle. In this section we would like to put forward a realisation of this mechanism in the context of 3D $\NN=2$ quivers deconstructing 4D $\NN=2$ SYM theories. We can easily compute the $S_b^3$ partition functions in the presence of these local modifications and deduce, after deconstruction, the form of the four-dimensional superconformal index including contributions that we will interpret as coming from an arbitrary number of codimension-one $\frac{1}{2}$-BPS defect insertions.

\begin{figure}
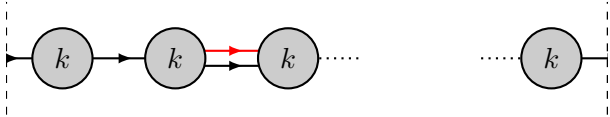
 
\begin{center} 
  \drawquivercodimonea
\end{center}
\caption{3D $\mathcal N=2$ circular quiver with $N$ nodes and a localised insertion. The black nodes denote $U(k)$ vector multiplets, while the black links bifundamental-chiral multiplets with assigned R charge $r=1$. The endpoints are to be understood as periodically identified. The red link is a bifundamental-chiral multiplet with generic R charge.}\label{fig:3} 
\end{figure}

Let us reconsider the 3D quiver of Fig.\ \ref{fig:2} that deconstructs the 4D $\NN=2$ SYM theory on $S^1\times \IR^3$. Each node represents a 3D $\NN=2$ vector multiplet with $U(k)$ gauge group and each  link a bifundamental chiral superfield. In Fig.\ \ref{fig:3} this quiver has been modified by adding a single extra bifundamental link denoted by a red arrow. In this case, we perform deconstruction by giving a vev to all the black bifundamentals while leaving the new extra bifundamental without a vev. One can check, at the level of the classical action, that the limit \eqref{eq:2}, \eqref{eq:1}, deconstructs the 4D $\NN=2$ SYM theory with a 3D defect localised on the emerging $S^1$. The 3D defect is described by a gauged 3D $\NN=2$ chiral superfield in the adjoint representation of the bulk $U(k)$ gauge group.

In this setup, and at the level of the $S^3_b$ partition function, exact deconstruction works in the following manner. Making use of the results of \eqref{simplequiver}, the circular quiver of Fig.\ \ref{fig:3} admits the partition function
\begin{align}
\begin{split}
\label{co1aa}
\mathcal Z^{\textrm{quiver}}_{\mathrm{3D}} &=\prod_{\alpha}\frac{1}{k!}\int \prod_{b=1}^{k} \mathrm d \sigma_b^{(\alpha)} \Delta^{\text{Haar}} \left(\sigma^{(\alpha)}\right) \prod_{b,c=1}^{k} \frac{\Gamma_h\left(\omega_+ + \sigma^{(\alpha)}_b - \sigma^{(\alpha+1)}_c \Big| \omega_1, \omega_2\right)}{ \widehat{\Gamma}_h\left(\sigma^{(\alpha)}_b - \sigma^{(\alpha)}_c \Big| \omega_1, \omega_2\right) }
\\
& \hspace{1cm}\times \prod_{b,c=1}^{k} \Gamma_h\left(\omega_+ r+ \sigma^{(0)}_b - \sigma^{(1)}_c \Big| \omega_1, \omega_2\right)~.
\end{split}
\end{align}
The first line includes the contributions to the partition function without the new insertion. The second line incorporates the insertion, which for concreteness is taken as a bifundamental link between the $0$th and the 1st node; the precise location will turn out not to be important. The $U(1)_R$ charge of the bifundamentals in the first line has been chosen as $r=1$, while that of the bifundamental insertion is left arbitrary.

Exact deconstruction is performed as in previous sections. On the first line we identify $\sigma^{(\alpha)}_b\to \sigma_b$  and shift the arguments of the $\Gamma_h$ functions by $\frac{\alpha}{R}$. Note, however, that while on the second line the $\sigma$'s are also identified, the argument of the $\Gamma_h$ function is not shifted since we do not give a vev to the corresponding bifundamental scalar. This leads to the partition function
\begin{align}
\begin{split}
\label{co1ab}
\mathcal Z^{\textrm{Dec}}_{\mathrm{3D}} &= \frac{1}{k!}\int \prod_{b=1}^{k}\mathrm d \sigma_b \prod_{\alpha}\frac{\Gamma_h(\omega_++ \frac{\alpha}{R}\Big |\omega_1, \omega_2)^{k}}{\widehat\Gamma_h(\frac{\alpha}{R}|\omega_1,\omega_2)^{k}} \prod_{b\neq c} \frac{\Gamma_h\left(\omega_+  + \sigma_b - \sigma_c  + \frac{\alpha}{R}\Big| \omega_1, \omega_2\right)}{ \Gamma_h\left(\sigma_b - \sigma_c + \frac{\alpha}{R}\Big| \omega_1, \omega_2\right) }\\
& \hspace{1cm}\times  \prod_{b,c=1}^{k} \Gamma_h\left(\omega_+ r+ \sigma_b - \sigma_c \Big| \omega_1, \omega_2\right)
\\
&=\frac{1}{k!} (\fp;\fp)^{k} (\fq;\fq)^{k}\oint \prod_{b=1}^{k}\frac{\mathrm d v_b}{2 \pi i v_b} \frac{\prod_{b,c=1}^{k} \Gamma_h\left(\frac{1}{2 \pi i R} \log[ (\fp\fq)^{\frac{r}{2}}v_c v_b^{-1}]\Big| \frac{\log\fp}{ 2\pi i R}, \frac{\log\fq}{ 2\pi i R}\right)}{\prod_{b\neq c}\Gamma_e\left(v_bv_c^{-1}\Big|\fp,\fq\right)}~,
\end{split}
\end{align}
where we have used the standard relations between 3D chemical potentials and 4D fugacities from Sec.~\ref{N2decsec}. This can be interpreted as the expression for the $\NN = 1$ index of a 4D $\NN=2$ vector multiplet in the presence of a codimension-one $\frac{1}{2}$-BPS defect. Exact deconstruction has converted the hyperbolic Gamma functions of the nodes and Higgsed bifundamentals into elliptic Gamma functions but has left behind a single hyperbolic Gamma function that represents the contribution of the gauged 3D defect.

\begin{figure}
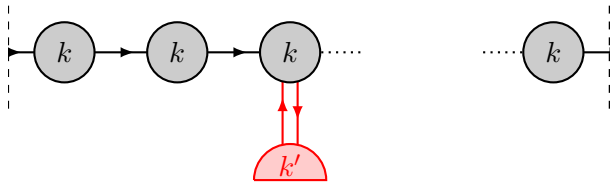
 
\begin{center} 
  \drawquivercodimoneb
\end{center}
\caption{3D $\mathcal N=2$ circular quiver with $N$ nodes and a different type of localised insertion. The black nodes denote $U(k)$ vector multiplets, while the black links bifundamental-chiral multiplets with assigned R charge $r=1$. The endpoints are to be understood as periodically identified. The red nodes are $U(k')$ vector multiplets while the red links bifundamental-chiral multiplets with generic R charge.}\label{fig:4} 
\end{figure}

Other types of localised modifications can be considered in a similar fashion. In Fig.\ \ref{fig:4} we add an extra 3D node with gauge group $U(k')$ linked to the 0th standard $U(k)$ node by a $U(k)\times U(k')$ bifundamental 3D hypermultiplet. In that case the $S_b^3$ partition function is
\begin{align}
\begin{split}
\label{co1ac}
\mathcal Z^{\textrm{quiver}}_{\mathrm{3D}} &=\prod_{\alpha}\frac{1}{k!}\int \prod_{b=1}^{k} \mathrm d \sigma_b^{(\alpha)} \Delta^{\text{Haar}} \left(\sigma^{(\alpha)}\right) \prod_{b,c=1}^{k} \frac{\Gamma_h\left(\omega_+ + \sigma^{(\alpha)}_b - \sigma^{(\alpha+1)}_c \Big| \omega_1, \omega_2\right)}{ \widehat{\Gamma}_h\left(\sigma^{(\alpha)}_b - \sigma^{(\alpha)}_c \Big| \omega_1, \omega_2\right) }
\\
& \hspace{0.8cm}\times \frac{1}{k'!}\int \prod_{\hat{b}=1}^{k'} \mathrm d \lambda_{\hat{b}} \Delta^{\text{Haar}} \left(\lambda\right) \frac{\prod_{\hat{b}=1}^{k}\prod_{\hat{c}=1}^{k'}\Gamma_h\left(\omega_+ r \pm (\sigma^{(0)}_b - \lambda_{\hat{c}})\Big| \omega_1, \omega_2\right)}{\prod_{\hat{b},\hat{c}=1}^{k'} \widehat{\Gamma}_h\left(\lambda_{\hat{b}} - \lambda_{\hat{c}} \Big| \omega_1, \omega_2\right) } ~,
\end{split}
\end{align}
where we have denoted the Coulomb-branch parameters in the extra $U(k')$ in terms of  $\lambda_{\hat{b}}$. After exact deconstruction we obtain the 4D-3D partition function
\begin{align}
\begin{split}
  \label{co1ad}
  \mathcal Z^{\textrm{Dec}}_{\mathrm{3D}} & =\frac{1}{k!} (\fp;\fp)^{k} (\fq;\fq)^{k}\oint \prod_{b=1}^{k}\frac{\mathrm d v_b}{2 \pi i v_b} \prod_{b\neq c}\Gamma_e\left(v_bv_c^{-1}\Big|\fp,\fq\right)^{-1} \\
 & \times  \frac{1}{k'!}\int \prod_{\hat{b}=1}^{k'} \mathrm d \lambda_{\hat{b}} \Delta^{\text{Haar}} \left(\lambda\right) \frac{\prod_{\hat{b}=1}^{k}\prod_{\hat{c}=1}^{k'}\Gamma_h\left(\frac{\log(\fp\fq)^{\frac{r}{2}}}{ 2\pi i R} \pm (\frac{\log v_b}{ 2\pi i R} - \lambda_{\hat{c}})\Big|\frac{\log\fp}{ 2\pi i R}, \frac{\log\fq}{ 2\pi i R}\right)}{\prod_{\hat{b},\hat{c}=1}^{k'} \widehat{\Gamma}_h\left(\lambda_{\hat{b}} - \lambda_{\hat{c}} \Big| \frac{\log\fp}{ 2\pi i R}, \frac{\log\fq}{ 2\pi i R}\right) } 
~.
\end{split}
\end{align}

More sophisticated examples---defined by more involved gauge theories---can obviously be obtained in this context, or in the context of other 3D$\to$4D deconstructions. The resultant 4D-3D deconstructed partition functions will be a combination of a 4D superconformal index coupled to squashed-$S^3$ partition functions. The coupling between the two is realised by gauging (part of) the defect global symmetry by the gauge group of the bulk 4D theory. 

It is useful to contrast this situation with the situation considered e.g.\ in \cite{Gang:2012ff}, where the codimension-one defects in 4D $\NN=2$ theories are wrapping the $S^1 \times S^2 \subset S^1\times S^3$ , i.e.\ they are codimension-one inside the $S^3$. In that case the 4D-3D index is a combination of a 4D superconformal index coupled to 3D superconformal indices.

As a final comment, we observe that we can equally simply add an arbitrary number of localised insertions along the 3D quivers. Keeping the distance between the insertions finite along the circular quiver produces codimension-one defects post deconstruction that are situated at different positions along the $S^1$. The exact deconstruction of the $S^3$ partition function in this arrangement proceeds in the same manner as above and leads to a 4D superconformal index with insertions of several 3D squashed-$S^3$ partition functions. The final expression is independent of the original position of the localised modifications. This is consistent with the interpretation of these objects as defects that preserve half of the supersymmetry of the original superconformal index. We propose that, from the point of view of quantum mechanics along the $S^1$, we are inserting supersymmetric operators in the Witten index. A standard argument shows that the latter does not depend on the position of the insertions. For completeness, let us consider this argument in some detail.

Let $Q$ and $Q^\dagger$ be two of the supercharges preserved by the superconformal index. They obey the anticommutation relation $2\{Q,Q^\dagger\}=\HH$, where $\HH$ is the sum of the conformal Hamiltonian $H$, that implements translations in the `temporal' $S^1$ direction of $S^1\times S^3$, and a combination of spacetime and R-symmetry generators whose details depend on the specifics of the definition of the index. For $\NN=2$ theories on the round three-sphere our definition \eqref{eq:8} implies that $\HH=H-2j_2-2R-r$, where $j_2$ is the Cartan generator of the second factor of the spacetime $SU(2)_1\times SU(2)_2$ of the $S^3$ and $R, r$ are the Cartan generators of the R-symmetry group $SU(2)_R\times U(1)_r$. Consider R-neutral local operators $\OO$ with $j_2=0$ annihilated by, say, $Q^\dagger$. Eventually, we would like to propose that our co-dimension-one defects are such operators.

The 4D index in the presence of co-dimension-one defects can be expressed in terms of $n$-point functions of $n$ such operator insertions\footnote{Note that in the absence of insertions this expression reverts back to the usual definition of the superconformal index \eqref{eq:8}.}
\beq
\label{co1ae}
\sum_i \langle \psi_i| (-1)^F\fp^{j_2 - j_1 - r}\fq^{j_2 +j_1 - r}\ft^{R + r} e^{(-\frac{\beta}{2}-\tau_1)\mathcal H}\OO_1 e^{(\tau_1-\tau_2)\mathcal H}\cdots e^{(\tau_{n-1}-\tau_{n})\mathcal H}\OO_n e^{(\tau_n-\frac{\beta}{2})\mathcal H}|\psi_i \rangle\;,
\eeq
where the sum is taken over the quantum-mechanical Hilbert space. This Hilbert space is $\mathbb Z_2$ graded by the supercharge $Q$ and, in the absence of $\mathcal O$ insertions, the non-ground states pairwise cancel as is usual for a Witten-type index. It is straightforward to check that the insertion of $\frac{1}{2}$-BPS operators $\mathcal O$ as in \eqref{co1ae} does not spoil this argument: For every non-ground state $|\alpha\rangle$ in the sum \eqref{co1ae} the state $Q^\dagger |\alpha \rangle$ once again contributes the same amount with an opposite sign leading to a pairwise cancellation. One is therefore left with the correlators involving ground states, which are however independent of the positions $\tau_i$. Consider for instance a sample ground state $| 0\rangle$. Shifting the position $\tau_i$ of the operator $\OO_i$ is equivalent to the action of $\HH$ on $\OO_i$, $[\HH, \OO_i]=2\{ Q^\dagger, [Q,\OO_i]\}$. As we move the supercharge $Q^\dagger$ over to the other $\frac{1}{2}$-BPS insertions inside the correlation function everything is annihilated by supersymmetry yielding 
\beq
\label{co1af}
\frac{\d}{\d\tau_i} \langle 0| (-1)^F\fp^{j_2 - j_1 - r}\fq^{j_2 +j_1 - r}\ft^{R + r} \OO_1(\tau_1) \cdots \OO_n(\tau_n) |0 \rangle=0
~.
\eeq

Returning to the results of deconstruction, we notice that the
obtained partition functions are similarly independent of the
positions of the deconstructed co-dimension-one defects. By
construction, these defects are supersymmetric, R- and $j_2$ neutral
and from the quantum mechanics point of view they are local operators
in the `time' direction. The fact that their $S^3\times S^1$ partition
function is $\tau_i$-independent agrees with their interpretation as
$\frac{1}{2}$-BPS defects in quantum mechanics and the argument given
in the previous paragraph.\footnote{It would be interesting to explore
  this connection in more detail. One would generally expect that the
  operators $\OO$ admit a realisation in terms of a sum $\OO =  c_{\OO}^{ij}| \psi_i \rangle \langle \psi_j|$, where the coefficients are given by $ c_{\OO}^{ij} \sim \langle \psi_i | \OO | \psi_j \rangle$. In particular, the coefficient $\langle 0 | \OO | 0\rangle$ should correspond to a pure $S^3$ partition function. It is through these coefficients that the logarithmic dependence in the fugacities that enters expressions such as E.q.~\eqref{co1ad} can be captured from the point of view of quantum mechanics.}

\section{Conclusions and Outlook}\label{conclusions}

In this paper we applied the exact-deconstruction procedure, introduced in \cite{Hayling:2017cva,Hayling:2018fmv}, to exact partition functions of 3D circular-quiver theories that lift to 4D indices. We generalised it to include supersymmetric, codimension-two defects that are associated with vortex- or Wilson-loop insertions at each node of the 3D quiver. In the process of doing so, we made use of some remarkable identities between special functions of hyperbolic and elliptic type. Even though we explicitly compared our post-deconstruction results with the index of superconformal 4D-2D systems, we stress that exact deconstruction more generally produces supersymmetric partition functions on $S^3\times S^1$ for  non-conformal setups. We note that we applied our procedure directly at the level of integrands for the 4D/3D and 2D/1D Coulomb-branch parameters. Although we have not done so, the associated integrals can subsequently be performed using the Jeffrey-Kirwan residue prescription \cite{Benini:2013xpa,Gadde:2013ftv,Assel:2015oxa}. It is also worth pointing out that by employing exact deconstruction we have straightforwardly recovered nontrivial results for 4D-2D indices while completely bypassing the conventional dimensional-deconstruction limit at the level of classical Lagrangians. Regarding the deconstruction of codimension-one defects, we introduced a localised insertion of gauge/matter fields at specific nodes of the 3D circular quiver, which lift to a coupled system of 4D-3D indices/three-sphere partition functions. Obtaining results for this class of defects is particularly simple through our method.

There are various avenues for future research stemming from this work. For example, it would be useful to further examine the prescription of Sec.~\ref{wilson} that isolates chiral-surface defect contributions in 4D related to a specific (single-column) Young tableau and, more specifically, contemplate further on its interpretation from the four-dimensional perspective. Moreover, vortex defects in both 4D and 3D can be obtained using certain difference operators that act directly on the 4D/3D index/partition function \cite{Gaiotto:2012xa,Bullimore:2014nla}. These operators satisfy an interesting elliptic algebra and are related by dimensional reduction. It would be interesting to see how exact deconstruction fits into this picture. In the direction of reducing supersymmetry, it would be worth determining whether the surface defects for 4D $\NN = 1$ SCFTs recently discussed in \cite{Razamat:2018zel} can also be studied from three dimensions using exact deconstruction. 

Departing from our 3D/4D setup, an obvious generalisation of our results would involve applying them to the six-dimensional (2,0) theory \cite{ArkaniHamed:2001ie,Hayling:2017cva}. One could attempt to deconstruct the (2,0) partition function on $S^4 \times T^2$ in the presence of various defects, based on the $S^4$ (defect) partition functions associated with $\NN = 2$ superconformal circular-quiver theories in 4D. Such an approach would make contact with and supplement the results of \cite{Bullimore:2014upa}.

More generally, exact deconstruction---not necessarily involving defects---could be applicable in a wide range of contexts. For instance, one could use it sequentially starting from the partition function of a 3D toroidal-quiver theory on $S^2\times S^1$ to first recover the $S^2 \times T^2$ partition function of a 4D circular quiver and then to get the 6D (2,0) partition function on $S^2\times T^4$. In addition, it could be implemented on the three-sphere partition function for an orbifold of ABJM theory to reproduce the results of \cite{Bourton:2018jwb} for the index of non-Lagrangian $\NN=3$ SCFTs \cite{Garcia-Etxebarria:2015wns,Aharony:2016kai}. Finally, it would be very interesting to try and extend the principle of deconstruction to non-Lagrangian starting points. This would entail generalising exact deconstruction away from the individual building-block prescription presented here, to an operation at the level of the full partition function. We hope to report on some of these items in the near future.


\ack{ \bigskip We would like to thank M.~Buican, M.~Bullimore, S.~Cremonesi, F.~Nieri, L.~Rastelli and P.~Richmond for helpful discussions and comments, as well as E.~Pomoni for collaboration at the initial stages of this project. J.H.\ is funded through an STFC research studentship and would like to thank CERN for hospitality during various phases of this work. V.N.\ would like to thank the Galileo Galilei Institute for Theoretical Physics for the hospitality and the INFN for partial support during the completion of this work. The work of V.N.\ is supported by STFC under the consolidated grant ST/P000371/1. C.P.\ is supported by the Royal Society through a University Research Fellowship, grant number UF120032.}

\begin{appendix}
  
\section{Useful Mathematical Definitions and Identities}
\label{math}

In this appendix we present a self-contained summary of the special functions that play a central r\^ole in the discussion of the main text. We highlight the mathematical identities that implement deconstruction from three-dimensional QFTs with one-dimensional defects to four-dimensional QFTs with two-dimensional defects.

A basic component of the functions that we consider in this paper is {\it Barnes' $N$-Gamma function}
\beq
\label{mathaa}
\Gamma_N (x | \omega_1, \ldots, \omega_N) \equiv \exp\left[\frac{\partial}{\partial s} \zeta_N (s,x |\omega_1, \ldots, \omega_N)\bigg|_{s=0}\right]
~,
\eeq
which is defined through the analytic continuation of the Barnes $N$-zeta function 
\beq
\label{mathab}
\zeta_N (s,x |\omega_1, \ldots, \omega_N)   = \sum_{\bm{\ell} \in \mathbb{N}^N} \left(x + \bm{\ell} \cdot \bm{\omega}\right)^{-s}
~.
\eeq
This function is convergent for $\mathfrak{R}\omega_i >0$. We will use the zeta-regularised infinite-product representation
\begin{align}\label{gamma}
\Gamma_N (x | \omega_1,\ldots, \omega_N) = \ZProd_{\bm{\ell} \in \mathbb N^N}\frac{1}{x + \bm{\ell} \cdot \bm{\omega}}\;.
\end{align}
We will also define a similar function, called $\widehat{\Gamma}_N$, whose infinite product representation is \eqref{gamma} with $\bm{\ell} \in \mathbb{N}^N \setminus \{\mathbf{0}\}$.

The {\it hyperbolic Gamma function} $\Gamma_h(x | \omega_1,\omega_2)$, introduced by Ruijsenaars \cite{Ruijsenaars:1997}, is closely related to the double-Gamma functions $\Gamma_2$ through the definition
\beq
\label{mathac}
\Gamma_h (x | \omega_1, \omega_2) \equiv \frac{\Gamma_2( -i x |-i \omega_1,-i \omega_2)}{\Gamma_2(i x - 2i \omega_+|-i \omega_1, -i \omega_2)}
= \prod_{\bm{\ell}\in \mathbb{N}^2 }\frac{-x   + (\ell_1+1) \omega_1 + (\ell_2+1) \omega_2}{x + \ell_1 \omega_1 + \ell_2 \omega_2} 
~,
\eeq
where
\beq
\label{mathad}
\omega_\pm := \frac{\omega_1 \pm \omega_2}{2} 
~.  
\eeq 
In this particular definition, convergence requires $\Im \omega_i >0$. The hyperbolic Gamma function---and its counterpart $\widehat{\Gamma}_h$, defined using the $\widehat{\Gamma}_N$s---appears prominently in the computation of the $S^3$ partition function of Lagrangian 3D $\NN\geq 2$ supersymmetric QFTs through supersymmetric localisation \cite{Kapustin:2009kz,Hama:2010av,Hama:2011ea}. In this context, the auxiliary parameters $\omega_1$, $\omega_2$ are $\omega_1 =ib$, $\omega_2 = ib^{-1}$, where the positive real number $b=\sqrt{\frac{\tilde \ell}{\ell}}$ encodes the squashing of $S^3$ into a $U(1)\times U(1)$ isometric hyper-ellipsoid (see also Eq.~\eqref{eq:4} in the main text).

The {\it elliptic Gamma function} $\Gamma_e(z | \fp, \fq)$ (for $|\fp|, |\fq|<1$) is also useful. It has the infinite-product representation
\beq
\label{mathae}
\Gamma_e (z | \fp, \fq) = \prod_{\bm{\ell}\in \mathbb{N}^2} \frac{1 - z^{-1} \fp^{\ell_1+1}\fq^{\ell_2+1}}{1-z\, \fp^{\ell_1} \fq^{\ell_2}}
~.
\eeq
This function appears in the computation of the superconformal index
of four-dimensional supersymmetric QFTs on $S_R^1 \times S^3$, where
$R$ denotes the (dimensionless) radius of the circle. The degeneration limit \cite{van2007hyperbolic,Ruijsenaars1987}
\beq
\label{mathaf}
\lim_{R\to 0^+} \Gamma_e\left( e^{2\pi i R x} | e^{2\pi i \omega_1 R}, e^{2\pi i \omega_2 R} \right) e^{\frac{\pi i(x-\omega_+)}{12 R\omega_1\omega_2}} = \Gamma_h(x | \omega_1,\omega_2)
\eeq
implements at the level of the superconformal index the small-radius limit of the $S^1$ that leads to the reduction from four to three dimensions \cite{Dolan:2011rp,Gadde:2011ia}.

In this paper we are interested in the opposite process where an operation on a 3D theory will deconstruct a 4D theory. The mathematical identity that implements this process is the infinite-product formula
\beq
\label{mathag}
\prod_{\alpha = -\infty}^\infty \Gamma_h\left(x + \frac{\alpha}{R}\bigg| \omega_1 , \omega_2\right) = {\mathfrak x}^2 \left(\fp \fq\right)^{-\frac{1}{2}} \Gamma_e (\mathfrak x |\fp, \fq)
\eeq
for $\mathfrak x=e^{2\pi i R x}$, $\fp = e^{2\pi i R \omega_1}$, $\fq=
e^{2\pi i R \omega_2}$. The combination $R\sqrt{\ell\tilde \ell}$
plays the r\^ole of the physical radius of the deconstructed $S^1$. The proof of this formula follows from a straightforward manipulation of the infinite-product representation of the hyperbolic Gamma functions, \eqref{mathac}, where the main ingredient is the infinite-product presentation of the sine function
\beq
\label{mathai}
\sin(\pi w) = \pi w \prod_{n=1}^\infty \left( 1- \frac{w^2}{n^2} \right)
~.
\eeq

The Gamma function $\Gamma_1$, which is proportional to the ordinary $\Gamma$ function,
\beq
\label{mathaj}
\Gamma_1( x|\omega ) = \frac{\omega^{\omega^{-1}x-\frac{1}{2}}}{\sqrt{2\pi}} \Gamma(x\omega^{-1})
~,
\eeq
appears when we discuss the partition functions of $\NN=2$ supersymmetric quantum mechanics on 1D defects of 3D QFTs. One can show that the following identity is satisfied
\beq
\label{mathak}
\Gamma_1(x |\omega)^{-1} \Gamma_1(\omega -x | \omega)^{-1} = 2 \sin\left( \frac{\pi x}{\omega} \right)
~.
\eeq
In partition functions we mainly encounter the following combination of $\Gamma_1$ functions
\beq
\label{mathal}
\Delta_h( a | \omega, t ) \equiv \frac{\Gamma_1 (a|\omega) \Gamma_1 (\omega - a|\omega)}
{\Gamma_1 (a+t |\omega) \Gamma_1 (\omega - a - t |\omega)}
= \frac{\sin(\frac{\pi(a+t)}{\omega})}{\sin(\frac{\pi a}{\omega})}
~.  
\eeq

The analogue of the deconstruction identity \eqref{mathag} for the function $\Delta_h$ is
\beq
\label{matham}
\Delta_e(A | \Omega,T) = \prod_{\alpha=-\infty}^\infty \Delta_h \left( a +i \frac{\alpha}{R} \Big | \omega, t \right)
~,
\eeq
where $A= e^{-2\pi  R a}$, $\Omega= e^{-2\pi   R \omega }$, $T = e^{-2\pi   R t}$. The function $\Delta_e(A | \Omega, T)$ is the ratio of theta functions
\beq
\label{mathan}
\Delta(A| \Omega, T) = \frac{\theta(A T  |\Omega)}{\theta(A | \Omega)}
~,
\eeq
where
\beq
\label{mathao}
\theta(z|\fq) = \prod_{n=0}^{\infty} (1-z \fq^n)(1-z^{-1} \fq^{n+1})
~.
\eeq
This particular ratio appears in 4D $\NN=2$ superconformal indices with insertions of codimension-two defects  \cite{Gaiotto:2012xa,Gadde:2013ftv}. The identity \eqref{matham} plays an instrumental r\^ole when we use 1D defects in 3D QFTs to deconstruct codimension-two surface operators in 4D QFTs.

\end{appendix}


\newpage

\bibliography{3D4Ddeconstruction}

\end{document}